\documentclass[sigconf, 10pt]{acmart}

\renewcommand\footnotetextcopyrightpermission[1]{}
\usepackage[subrefformat=parens, labelformat=parens]{subfig}
\usepackage[font=small]{caption}
\usepackage[ruled, vlined, linesnumbered]{algorithm2e}
\usepackage{amsfonts}
\usepackage{amsmath}

\usepackage{amssymb}
\usepackage{amsthm}
\usepackage{gensymb}
\usepackage{soul}
\usepackage{graphicx}
\usepackage{enumitem}
\usepackage{booktabs}
\usepackage{rotating}
\usepackage{multirow}

\usepackage{hyperref}
\usepackage{subcaption}
\usepackage{array}
\usepackage{siunitx}
\usepackage{tcolorbox}
\usepackage{tikz}
\usepackage[dvipsnames]{xcolor}

\begin{document}

\setlength{\abovedisplayskip}{6pt}
\setlength{\belowdisplayskip}{6pt}

\thispagestyle{plain}
\pagestyle{plain}

%% COMMENTS
\definecolor{lightgray}{gray}{0.9}
\definecolor{lightblue}{rgb}{0.9,0.9,1}
\definecolor{LightMagenta}{rgb}{1,0.5,1}
\definecolor{red}{rgb}{1,0,0}

\newcommand\couldremove[1]{{\color{lightgray} #1}}
\newcommand{\remove}[1]{}
\newcommand{\move}[2]{ {\textcolor{Purple}{ \bf --- MOVE #1: --- }} {\textcolor{Orchid}{#2}} }

%% FOR COMMENTS & EDITS
\newcommand{\hlc}[2][yellow]{ {\sethlcolor{#1} \hl{#2}} }
\newcommand\note[1]{\hlc[SkyBlue]{-- #1 --}} % highlighted notes of other colors.
% For colors info from xcolor package, check out:
% http://en.wikibooks.org/wiki/LaTeX/Colors

\newcommand\mynote[1]{\hlc[yellow]{#1}}
\newcommand\zhihui[1]{\hlc[BurntOrange]{ZG: #1}}
\newcommand\zhecun[1]{\hlc[green]{ZL: #1}}
\newcommand\tingjun[1]{\hlc[yellow]{TC: #1}}

\newcommand\name{{\sc Chameleon}}
\newcommand\namebf{{\sc\textbf{Chameleon}}}

\newcommand{\myparatight}[1]{\vspace{0.5ex}\noindent\textbf{#1~~}}
\newcommand*\circled[1]{\tikz[baseline=(char.base)]{
            \node[shape=circle,draw,fill=black,text=white,inner sep=0.6pt] (char) {\scriptsize{#1}};}}

\newcommand{\iu}{{j}}
\newcommand{\eu}{{e}}
\newcommand\usec{$\upmu$s}
\newcommand\usecbf{$\boldsymbol{\upmu}$s}

\newcommand{\littlesum}{\mathop{\textstyle\sum}}

\newcommand{\complexity}[1]{\mathcal{O}(#1)}

\newcommand{\myAbs}[1]{\left|{#1}\right|}
\newcommand{\myNorm}[1]{\left|\left|{#1}\right|\right|}
\newcommand{\myAng}[1]{\angle{#1}}
\newcommand{\myConj}[1]{{#1}^{*}}
\newcommand{\myTrans}[1]{{#1}^{\top}}

\newcommand{\waveTx}{x}
\newcommand{\waveTxVec}{\mathbf{x}}
\newcommand{\specTx}{X}
\newcommand{\specTxVec}{\mathbf{X}}
\newcommand{\waveRx}{y}
\newcommand{\waveRxVec}{\mathbf{y}}
\newcommand{\specRx}{Y}
\newcommand{\specRxVec}{\mathbf{Y}}
\newcommand{\waveIdx}{n}
\newcommand{\specIdx}{k}
\newcommand{\waveNoise}{\mathcal{N}}
\newcommand{\pathDelay}{\tau}
\newcommand{\pathAmp}{\alpha}
\newcommand{\pathPhase}{\Delta\theta}
\newcommand{\sampRate}{f_s}
\newcommand{\CSI}{H}
\newcommand{\CSIVec}{\mathbf{H}}
\newcommand{\fftSize}{N_{\textrm{fft}}}
\newcommand{\fft}{\textsf{FFT}}
\newcommand{\ifft}{\textsf{FFT}^{-1}}

\newcommand{\antNum}{N}
\newcommand{\antIdx}{n}
\newcommand{\antDist}{d}
\newcommand{\steerVec}{\mathbf{s}}
\newcommand{\steer}{s}
\newcommand{\angleAz}{\phi}
\newcommand{\angleEl}{\psi}
\newcommand{\waveLength}{\lambda}
\newcommand{\bfWeight}{w}
\newcommand{\bfWeightVec}{\mathbf{w}}
\newcommand{\bfWeightMat}{\mathbf{W}}
\newcommand{\bfGain}{g}
\newcommand{\bfGainAll}{G}

\newcommand{\userNum}{U}
\newcommand{\userIdx}{u}
\newcommand{\angleUserAz}[1]{\phi_{#1}^{(c)}}
\newcommand{\angleUserNew}[1]{\widetilde{\phi}_{#1}^{(c)}}
\newcommand{\angleUserEl}[1]{\psi_{#1}^{(c)}}
\newcommand{\angleSenseAz}{\phi^{(s)}}
\newcommand{\angleSenseEl}{\psi^{(s)}}
\newcommand{\angleSenseSet}{\boldsymbol{\Phi}^{(s)}}
\newcommand{\angleSenseNum}{M}
\newcommand{\angleSenseIdx}{m}

\newcommand{\bfGainComm}[1]{g^{(c)}_{{#1}}}
\newcommand{\bfGainCommMean}[1]{g^{(c)}_{{#1}}}
\newcommand{\snrBaseComm}[1]{\gamma^{(c)}_{{#1}}}
\newcommand{\snrComm}[1]{\Gamma^{(c)}_{{#1}}}
\newcommand{\snrCommNew}[1]{\widetilde{\Gamma}^{(c)}_{{#1}}}
\newcommand{\speedComm}[1]{R^{(c)}_{{#1}}}
\newcommand{\bfGainSense}{g^{(s)}}
\newcommand{\snrBaseSense}{\gamma^{(s)}}
\newcommand{\snrSense}{\Gamma^{(s)}}
\newcommand{\speedSense}{R^{(s)}}
\newcommand{\bfWeightVecSet}{\mathcal{W}}
\newcommand{\snrCommMin}{\Gamma^{\textrm{min}}}
\newcommand{\snrCommMinSet}{\boldsymbol{\Gamma}_{\textrm{min}}}

\newcommand{\bfGainData}[1]{g^{(d)}_{{#1}}}
\newcommand{\bfGainDataMean}{g^{(d)}}
\newcommand{\bfGainGap}[1]{\Delta g_{{#1}}}
\newcommand{\waveTxCal}{\Tilde{x}}
\newcommand{\waveTxVecCal}{\Tilde{\mathbf{x}}}

\newcommand{\snr}[1]{\gamma_{#1}}
\newcommand{\sinr}[1]{\Gamma_{#1}}
\newcommand{\speed}[1]{R_{#1}}
\newcommand{\bandwidth}{B}
\newcommand{\bfGainSenseMin}{g^{c}_{\textrm{min}}}
\newcommand{\bfGainSenseOpt}{g^{s\star}}
\newcommand{\bfWeightOpt}{\mathbf{w}^{\star}}
\newcommand{\perturb}{\epsilon}

\newcommand{\BeamOptimOrig}{\textsf{OPT-Base}}
\newcommand{\BeamOptim}{\textsf{OPT-Accel}}
\newcommand{\DelayOptim}{\textsf{OPT-Delay}}

\newcommand{\waveTxLow}[1]{x_{#1}}
\newcommand{\waveTxLowVec}[1]{\mathbf{x}_{#1}}
\newcommand{\waveRxLow}[1]{y_{#1}}
\newcommand{\waveRxLowVec}[1]{\mathbf{y}_{#1}}
\newcommand{\specTxLow}[1]{X_{#1}}
\newcommand{\specTxLowVec}[1]{\mathbf{X}_{#1}}
\newcommand{\specTxLowCal}[1]{\Tilde{X}_{#1}}
\newcommand{\specTxLowVecCal}[1]{\Tilde{\mathbf{X}}_{#1}}
\newcommand{\specRxLow}[1]{Y_{#1}}
\newcommand{\specRxLowVec}[1]{\mathbf{Y}_{#1}}
\newcommand{\fftSizeLow}{N_{\textrm{fft}}^{\prime}}
\newcommand{\waveDelay}{\Delta \waveIdx}
\newcommand{\waveDelayNum}{N_{\Delta \waveIdx}}
\newcommand{\CSILow}[1]{H^{(s)}_{#1}}
\newcommand{\CSILowVec}[1]{\mathbf{H}^{(s)}_{#1}}
\newcommand{\loss}{L}
\newcommand{\phaseWeight}{\nu}
\newcommand{\phaseWeightVec}{\boldsymbol{\nu}}
\newcommand{\phaseSlope}{\zeta}
\newcommand{\phaseBias}{\xi}
\newcommand{\waveDelayOpt}{\Delta \waveIdx^{\star}}
\newcommand{\CSILowOpt}[1]{H^{(s)\star}_{#1}}
\newcommand{\CSILowVecOpt}[1]{\mathbf{H}^{(s)\star}_{#1}}

\newcommand{\waveRxLowCal}[1]{\widetilde{y}_{#1}}
\newcommand{\waveRxLowCalVec}[1]{\widetilde{\mathbf{y}}_{#1}}
\newcommand{\CSICommVec}[1]{\mathbf{H}^{(c)}_{#1}}
\newcommand{\CSIComm}[1]{H^{(c)}_{#1}}

\title{{\namebf}: Integrated Sensing and Communication with Sub-Symbol Beam Switching in \\ mmWave Networks}

\author{Zhihui Gao, Zhecun Liu, Tingjun Chen}
\affiliation{%
  \institution{\vspace{0.5ex} Department of Electrical and Computer Engineering, Duke University\vspace{0.5ex}}
  \country{}
}
\email{{zhihui.gao, zhecun.liu, tingjun.chen}@duke.edu}

\begin{abstract}
Next-generation cellular networks are envisioned to integrate sensing capabilities with communication, particularly in the millimeter-wave (mmWave) spectrum, where beamforming using large-scale antenna arrays enables directional signal transmissions for improved spatial multiplexing.
In current 5G networks, however, beamforming is typically designed either for communication or sensing (e.g., beam training during link establishment).
In this paper, we present {\name}, a novel framework that augments and rapidly switches beamformers during each demodulation reference signal (DMRS) symbol to achieve integrated sensing and communication (ISAC) in 5G mmWave networks.
Each beamformer introduces an additional sensing beam toward target angles while maintaining the communication beams toward multiple users.
We implement {\name} on a {28}\thinspace{GHz} software-defined radio testbed supporting over-the-air 5G physical downlink shared channel (PDSCH) transmissions.
Extensive experiments in open environments show that {\name} achieves multi-user communication with a sum data rate of up to {0.80}\thinspace{Gbps} across two users.
Simultaneously, {\name} employs a beamformer switching interval of only {0.24}\thinspace{\usec}, therefore producing a {31}$\times${31}-point 2D imaging within just {0.875}\thinspace{ms}. 
Leveraging machine learning, {\name} further enables object localization with median errors of {0.14}\thinspace{m} (distance) and {0.24}$^{\circ}$ (angle), and material classification with 99.0\% accuracy.
\end{abstract}

\maketitle
\section{Introduction}
\label{sec: introduction}

%% figure begins
\begin{figure}%[!t]
    \centering
    % \vspace{-3mm}
    \includegraphics[width=0.99\columnwidth]{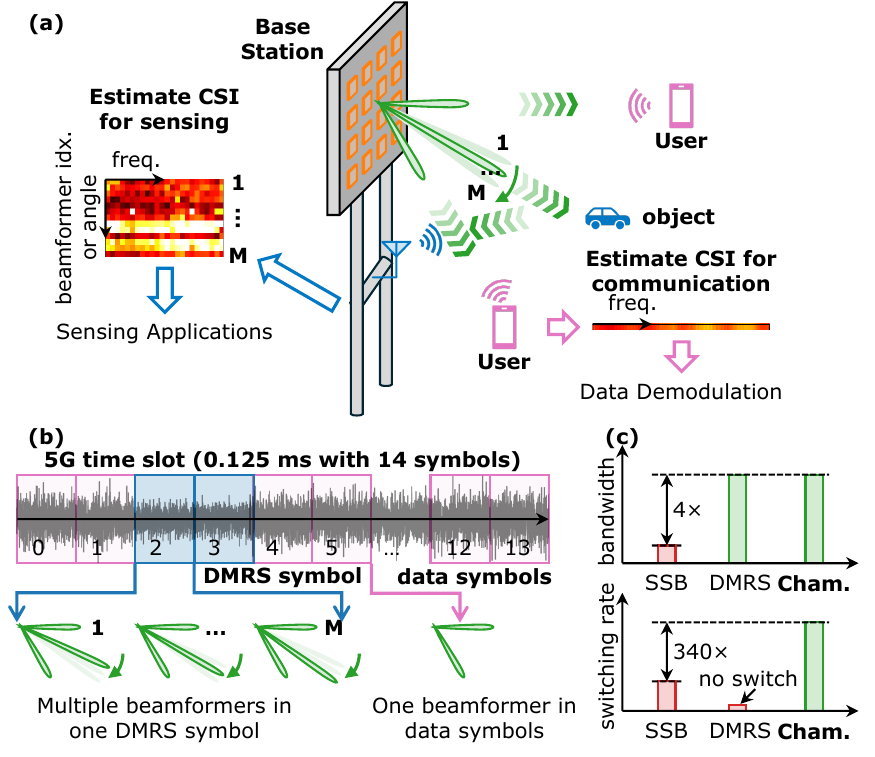}
    \vspace{-3mm}
    \caption{(a) In {\namebf}, the base station (BS) employs ISAC-capable beamformers toward multiple users for communication alongside one additional sensing beam that can be rapidly switched.
    (b) The sub-symbol beamformer during a single DMRS symbol is switched across multiple sensing angles, while maintaining the beams toward users; the beamformer during the data symbols is fixed, which is designed following the convention.
    (c) {\namebf} achieves {4}$\times$ larger bandwidth and $>${340}$\times$ faster beamformer switching rate than existing SSB-based approaches~\cite{abratkiewicz2023ssb, paidimarri2024eye, gallyas2024angle, pegoraro2024hisac}.
    }
    \vspace{-5mm}
    \label{fig:introduction-overview}
\end{figure}
%% figure ends

Next-generation cellular networks embrace integrated sensing and communication (ISAC) technology to seamlessly embed sensing capabilities into existing infrastructure to support a wide range of applications, such as autonomous vehicles~\cite{paidimarri2024eye, ding2023push, wang2020demystifying, ozkaptan2023mmwave, Nordio2025, Zhu2023}, super-resolution sensing for situational awareness~\cite{qian20203d, pegoraro2024hisac, yang2023slnet, zhang2020mmeye, gan2023poster}, and human-involved applications~\cite{xu2023work, zhang2024single, li2021vocalprint, ozturk2023radio, Liu2024}.
This trend becomes more promising and feasible with the development of millimeter-wave (mmWave) networks, where beamforming is greatly involved to compensate for the high signal attenuation over distance at high frequencies.
The shorter wavelength of mmWave signals allows more antennas within the same antenna array, which potentially strengthens beamforming's sensing capability in two aspects: (\emph{i}) a higher beamforming gain on the reflected signal from the environment, and (\emph{ii}) a narrower beamwidth for improved angular resolution.

Despite the unique advantages of mmWave technology, its potential for ISAC remains underexplored in 5G networks. 
Most existing works~\cite{guan20213, abratkiewicz2023ssb, paidimarri2024eye, gallyas2024angle, pegoraro2024jump, pegoraro2024hisac} decouple the communication and sensing functionalities by allocating separate resources, e.g., in the time or frequency domain.
The beam training and tracking in the 5G signal synchronization signal block (SSB) steers a beam around the environment to search for users, where the reflected signals embed potential environmental information for sensing~\cite{abratkiewicz2023ssb, paidimarri2024eye, gallyas2024angle, pegoraro2024hisac}.
However, this process lacks simultaneous communication and is sparsely allocated with resource elements, e.g., occupying {28.8}\thinspace{MHz} for a {100}\thinspace{MHz} channel, and appearing {64} symbols every {20}\thinspace{ms}~\cite{sharetechnote, paidimarri2024eye}.
On the other hand, the physical downlink shared channel (PDSCH) that carries downlink (DL) data payload is used extensively for communication, whose demodulation reference symbols (DMRS) provide fine-grained channel state information (CSI) across the occupied signal bandwidth, and can be repurposed for sensing~\cite{ruan2022ipos}.
However, the beamformer over the whole PDSCH is fixed and steers the beams toward the users only, so the CSI lacks the flexibility to explore arbitrary areas of interest.
These facts motivate a new design for ISAC-capable beamformers with multiple beams directed to users for communication and an additional beam swept across the area of interest for sensing.

In this paper, we introduce {\name}\footnote{Unlike most animals, a chameleon can move each eye independently: one eye can focus on a specific object (\emph{communication}) while the other scans the surroundings (\emph{sensing}), offering a nearly {360}$^{\circ}$ field of vision.}, a novel ISAC framework tailored for 5G mmWave networks.
We design {\name} based on two key insights.
First, {\name} repurposes the DMRS symbols to serve dual purposes: the users estimate their \emph{communication CSI} for demodulating the data symbols for communication, while the base station (BS) estimates the \emph{sensing CSI} from the round-trip reflected signals for sensing, as shown in Fig.~\ref{fig:introduction-overview}(a).
Second, {\name} performs multiple beam switching during a single DMRS symbol (i.e., \emph{sub-symbol beam switching}), each toward a different sensing angle, as shown in Fig.~\ref{fig:introduction-overview}(b). 
The sensing beams can be swept at {340}$\times$ faster rate than existing works leveraging the SSBs~\cite{paidimarri2024eye} ({10}$\times$ via dense DMRS allocation and {34}$\times$ via sub-symbol switching), enhancing the time and angular resolution; in addition, the bandwidth of DMRS spans over entire transmission bandwidth, i.e., typically {4}$\times$ larger than that of the SSB, improving {\name}'s propagation delay perception with fine-grained CSI, as shown in Fig.~\ref{fig:introduction-overview}(c).

{\name} maintains communication beams toward the user(s) with slightly reduced power while repurposing the beamforming degrees-of-freedom (DoF) to create an additional sensing beam with sufficient power that can be rapidly switched across a set of desired sensing angles.
On the BS side, a receiver with a single data stream is used to capture the DMRS symbols reflected from the surrounding environment corresponding to each beamformer, which is then processed to extract the sensing CSI that can be used for various downstream tasks.
These tasks include, but are not limited to, distance (or time-of-flight) and angle estimations, as well as object classification and tracking.

The design of {\name} incorporates three key modules that function collaboratively to achieve the abovementioned goals:
The ISAC beamforming codebook design module derives the beamformer codebook for sub-symbol switching; the DMRS waveform pre-distortion module pre-compensates the transmitting DMRS waveform on the BS so that the users can effectively estimate the communication CSI for data demodulation; and the sensing CSI estimation module efficiently estimates the sensing CSI per beamformer on the BS.

We implement and evaluate {\name} using {28}\thinspace{GHz} software-defined radios (SDRs) deployed in the PAWR COSMOS testbed~\cite{raychaudhuri2020challenge}.
The BS includes two IBM {28}\thinspace{GHz} 64-element phased array antenna module (PAAM) boards~\cite{gu2021development, chen2023open} as the transmitter and the receiver, and a USRP N310 SDR~\cite{usrp_n310}.
A 5G PDSCH data link is established for a {92.16}\thinspace{MHz} link with a baseband sampling rate of {122.88}\thinspace{MHz}.
On the BS side, {\name} can estimate the effective sensing CSI of the reflected signals from a desired sensing angle per beamformer for sensing; on the user side, the communication CSI for demodulating the data symbols can still be estimated, which achieves up to {0.799}\thinspace{Gbps} sum data rate over two users.
The collected sensing CSI facilitates various downstream sensing applications. 
First, a 2D image of {31}$\times${31} pixels can be obtained by switching {31}$\times${31} beamformers with their respective sensing angles in azimuth and elevation within {0.875}\thinspace{ms}.
Then in assistance with machine learning (ML), the sensing CSI over {31} switching sub-symbol beamformers within a single DMRS symbol, i.e., a time duration of {8.3}\thinspace{\usec}, achieves a median distance error of {0.14}\thinspace{m} and a median angle error of {0.24}$^{\circ}$ for an object localization task, and a classification accuracy of {99\%} for a material classification task.

To summarize, the contributions of this paper include:
\begin{itemize}[leftmargin=*, topsep=2pt, itemsep=1pt]
    \item Integration of sub-symbol beam switching-based ISAC leveraging the dense DMRS symbols while sustaining multi-user communication in 5G mmWave networks;
    \item A novel CSI estimation algorithm for both signal demodulation and sensing, specifically designed for sub-symbol beam switching;
    \item Extensive experiments on a {28}\thinspace{GHz} SDR testbed, demonstrating the practicality and effectiveness of {\name} in real-world scenarios.
\end{itemize}

\noindent
\emph{To the best of our knowledge, this is the first work that truly integrates capability with 5G FR2 leveraging the dense DMRS symbols with sub-symbol beam switching.}
\section{Related Work}
\label{sec:related}
\myparatight{Novel mmWave sensing applications.}
In recent years, there has been growing interest in mmWave-based sensing for its short wavelength, along with its fine-grained sensing capability.
For example, mmWave pushes the resolution of the wireless imaging to the millimeter level~\cite{yang2023slnet, zhang2020mmeye, ding2023push, dodds2024around, Guan2020} or even comparable to point clouds generated by LiDARs~\cite{qian20203d, gan2023poster}.
Furthermore, the vibration within millimeters can also be captured by mmWave signals, which enables material classification for the different vibration features~\cite{wu2020msense, shanbhag2023contactless}.
In addition, human-involved sensing applications, such as gesture recognition~\cite{xu2023work, Akbar2023}, healthcare~\cite{zhang2024single, Zhang2023a, Shi2025, Liu2024}, and authentication~\cite{li2021vocalprint, ozturk2023radio, Wang2022}, can be enabled using mmWave signals.
However, most of these works rely on dedicated mmWave devices, e.g., mmWave radar, lacking simultaneous communication capability.

\myparatight{ISAC frameworks.}
Next-generation networks favor integrating sensing capability along with the communication.
Generally, ISAC can be categorized by its protocol, including Wi-Fi~\cite{schulz2018shadow, pegoraro2024jump, gu2021tyrloc, zhang2021widar3, hou2024rfboost}, 5G~\cite{abratkiewicz2023ssb, paidimarri2024eye, gallyas2024angle, pegoraro2024hisac, barneto2021full, ruan2022ipos, Ghoshal2023}, LoRa~\cite{Zhang2021}, and customized waveforms~\cite{guan20213, liu2020joint, Qian2025}.
Specifically for 5G mmWave networks, SSB~\cite{abratkiewicz2023ssb, paidimarri2024eye, gallyas2024angle, pegoraro2024hisac} is usually exploited for beam-switching-based ISAC, whose sensing rate and accessible bandwidth are still limited by the usage of SSB.
Close to our work, \cite{gallyas2024angle} also performs sub-symbol beam switching for object angle estimation, where only the received power is extracted and the communication performance is not explored.

\myparatight{Advanced beamforming on mmWave platforms.}
Emerging mmWave platforms with versatile beamforming capabilities contribute to efficient and stable communication.
For example, M5~\cite{zhang2022m5} generates multi-beams toward multi-users for broadcasting communications; mmReliable~\cite{jain2021two} generates two beams to avoid the blockage issue; Mambas~\cite{gao2024mambas} and Nulli-Fi~\cite{madani2021practical} utilize spatial nulling to mitigate the inter-user interference; Hawkeye~\cite{jian2019poster} considers a movable AP to enhance the LOS connectivity to the STA.
Moreover, LiDAR~\cite{woodford2021spacebeam} and mmWave radar~\cite{li2024map} are further employed for beam management.
In contrast, {\name} tailors beamforming to enable seamlessly integrated sensing capability while sustaining the desired communication performance in 5G mmWave networks.
\section{Preliminaries}
\label{sec:preliminaries}

\subsection{5G DMRS Symbols and CSI Estimation}
\label{ssec: preliminaries-dmrs-csi}

\myparatight{5G PDSCH Structure.}
In 5G, the physical downlink shared channel (PDSCH)~\cite{MATLAB5GToolbox, sharetechnote} carries downlink data between the BS and users.
There are two types of symbols in each time slot of the PDSCH: (\emph{i}) data symbols, where DL data is modulated; (\emph{ii}) demodulation reference signal (DMRS) symbols, where per-subcarrier channel state information (CSI) is estimated by users to demodulate the received data symbols. Among the {14} symbols within one time slot, there can be up to four DMRS symbols, and the rest are the data symbols.
In multi-user DL transmission, with $\userNum$ total users, the BS generates $\userNum$ baseband waveforms in the time domain, whose data symbols vary over users, while the DMRS symbols are pre-defined and identical across users.

\myparatight{DMRS-based CSI estimation.}
Assuming an FFT size of $\fftSize$, the time-domain waveform for each DMRS symbol consists of $\fftSize$ I/Q samples, excluding the cyclic prefix.
Let $\waveTxVec = [\waveTx[\waveIdx]] \in \mathbb{C}^{\fftSize}$ and $\waveRxVec = [\waveRx[\waveIdx]] \in \mathbb{C}^{\fftSize}$ denote the transmitted and received DMRS time-domain waveforms at the BS and users, respectively.
Applying an $\fftSize$-point FFT, the corresponding frequency-domain DMRS symbol is given by $\specTxVec = [\specTx[\specIdx]] \in \mathbb{C}^{\fftSize}$ and $\specRxVec = [\specRx[\specIdx]] \in \mathbb{C}^{\fftSize}$.
In 5G, the BS beamformer remains fixed within one PDSCH time slot over the {14} data/DMRS symbols, and the CSI associated with the beamformer can be extracted from any of the DMRS symbols within that slot. 
Specifically, the CSI vector $\CSIVec = [\CSI[\specIdx]] \in \mathbb{C}^{\fftSize}$ is estimated on a per-subcarrier basis by performing an element-wise division of the received and transmitted DMRS symbol in the frequency domain, i.e., $\CSI[\specIdx] = \specRx[\specIdx] / \specTx[\specIdx],\ \forall k$.

\myparatight{mmWave channel model.}
In the mmWave band, the over-the-air channel is usually dominated by the strongest path due to two primary reasons~\cite{paidimarri2024eye}:
(\emph{i}) mmWave signals experience severe attenuation with each reflection, causing multi-path components with very low signal power, and 
(\emph{ii}) directional beamforming focuses transmit power within a narrow angular range, making only paths aligned within that angular range significant.
The dominated path can be characterized by its amplitude attenuation $\pathAmp \in [0, 1]$, phase shift $\pathPhase \in [-\pi, +\pi)$, and propagation delay $\pathDelay \in [0, +\infty)$.
Given the transmission of DMRS symbol $\waveTx[\waveIdx]$, the received DMRS symbol $\waveRx[\waveIdx]$ can be approximated as
\begin{align}
    \waveRx[\waveIdx] \approx \pathAmp \eu^{\iu \pathPhase} \cdot \waveTx[\waveIdx-\sampRate\cdot\pathDelay] + \waveNoise[\waveIdx],
    \label{eq: channel-model}
\end{align}
where $\waveNoise[\waveIdx]$ denotes additive white Gaussian noise (AWGN), and $\sampRate$ is the receiver's sampling rate.
Under this model, the CSI exhibits a linear phase profile across subcarriers, with the slope corresponding to the propagation delay $\pathDelay$.
After clipping them into the range of $[-\pi, +\pi)$, the linear phases exhibit a \emph{sawtooth} shape.

\subsection{Beamforming on Antenna Array}

\myparatight{Antenna array geometry and steering vector.}
Without loss of generality, we consider a uniform linear array (ULA) composed of $\antNum$ antennas with an inter-antenna spacing of $\antDist$. 
All antenna elements are homogeneous, i.e., with the same maximum radiating power within their field of view (FoV).
Let $\steerVec(\angleAz) = [\steer_{\antIdx}(\angleAz)] \in \mathbb{C}^{\antNum}$ be the steering vector of the antenna array towards the azimuth angle $\angleAz$, it holds that
\begin{align}
    \steerVec(\angleAz) = \left[ 1, e^{\iu\frac{2\pi \antDist}{\waveLength} \sin\angleAz}, \dots, e^{\iu\frac{2\pi (\antNum-1) \antDist}{\waveLength} \sin\angleAz} \right],
\end{align}
where $\waveLength$ is the signal wavelength.
{\name} is also compatible with other antenna array geometries (e.g., planar or hexagonal arrays) by plugging in their steering vectors.

\myparatight{Beamforming weights and beamforming gain.}
We define $\bfWeightVec = [\bfWeight_{\antIdx}] \in \mathbb{C}^{\antNum}$ as the beamforming vector (a.k.a., \emph{beamformer}) for an $\antNum$-element antenna array; where $\myAbs{\bfWeight_{\antIdx}} \in [0, 1]$ and $\myAng{\bfWeight_{\antIdx}} \in [-\pi, +\pi)$ denote the amplitude and phase control of the $\antIdx$-th antenna.
The beamforming gain in the spatial direction of $\angleAz$, denoted by $\bfGain(\bfWeightVec, \angleAz)$, is given by
\begin{equation}
    \bfGain(\bfWeightVec, \angleAz) = \myAbs{\myTrans{\steerVec}(\angleAz) \cdot \bfWeightVec}^2.
\end{equation}
It is easy to see that a maximum beamforming gain of $\antNum^2$ can be achieved via conjugate beamforming, i.e., $\bfWeightVec = \myConj{\steerVec}(\angleAz)$.

\myparatight{Multi-user communication model.}
We consider a multi-user communication scenario between one BS and $\userNum$ users, each with a single data stream.
The dominant path (either LOS or NLOS) for the $\userIdx$-th user is denoted as the spatial angle $\angleUserAz{\userIdx}$ from the BS's perspective. 
Based on {\eqref{eq: channel-model}}, the received signal power at the $\userIdx$-th user can be determined by the strongest path at angle $\angleUserAz{\userIdx}$~\cite{gao2024mambas}.
Let $\snrBaseComm{\userIdx}$ denote the baseline communication (\emph{c}) SNR of the $\userIdx$-th user without beamforming.
When beamforming is applied, the effective SNR, $\snrComm{\userIdx}$, is given by
\begin{equation}
    \snrComm{\userIdx}(\bfWeightVec) = \snrBaseComm{\userIdx} \cdot \bfGain \left(\bfWeightVec, \angleUserAz{\userIdx} \right) = \snrBaseComm{\userIdx} \cdot \myAbs{\myTrans{\steerVec} \left(\angleUserAz{\userIdx} \right) \cdot \bfWeightVec}^2.
    \label{eq: communication-snr}
\end{equation}
In general, a higher SNR on the data symbols enables increased data rates, while a higher SNR on the DMRS symbols improves the accuracy of CSI estimation.

\myparatight{Monostatic sensing.}
{\name} is designed for monostatic sensing at the BS, where the BS is augmented with an auxiliary RX antenna (or array) to capture the reflected signals from the environment.
We assume that a fixed beamformer is applied at the auxiliary RX chain, which is synchronized with the BS transmitter.
To sense a target angle $\angleSenseAz$, a beamformer achieving high beamforming gain $\bfGain(\bfWeightVec, \angleSenseAz)$ is designed to enhance the round-trip signal reflections in that angle.
Let $\snrBaseSense$ denote the baseline SNR of the received signal for sensing (\emph{s}) without beamforming.
When sensing beamforming is applied, the effective SNR, $\snrSense$, is given by
\begin{equation}
    \snrSense(\bfWeightVec) = \snrBaseSense \cdot \bfGain \left(\bfWeightVec, \angleSenseAz \right) = \snrBaseSense \cdot \myAbs{\myTrans{\steerVec} \left(\angleSenseAz \right) \cdot \bfWeightVec}^2.
    \label{eq: sensing-snr}
\end{equation}

In summary, the beamforming gain (or SNR), either for communication (or $\snrComm{\userIdx}(\bfWeightVec)$) or sensing (or $\snrSense(\bfWeightVec)$), indicates the generated beam strength of the beamformer given by $\bfWeightVec$. To achieve ISAC for {\name}, both types of beamforming gains (or SNRs) should be maximized.
\section{System Design}

%% figure begins
\begin{figure}%[!t]
    \centering
    \includegraphics[width=0.99\columnwidth]{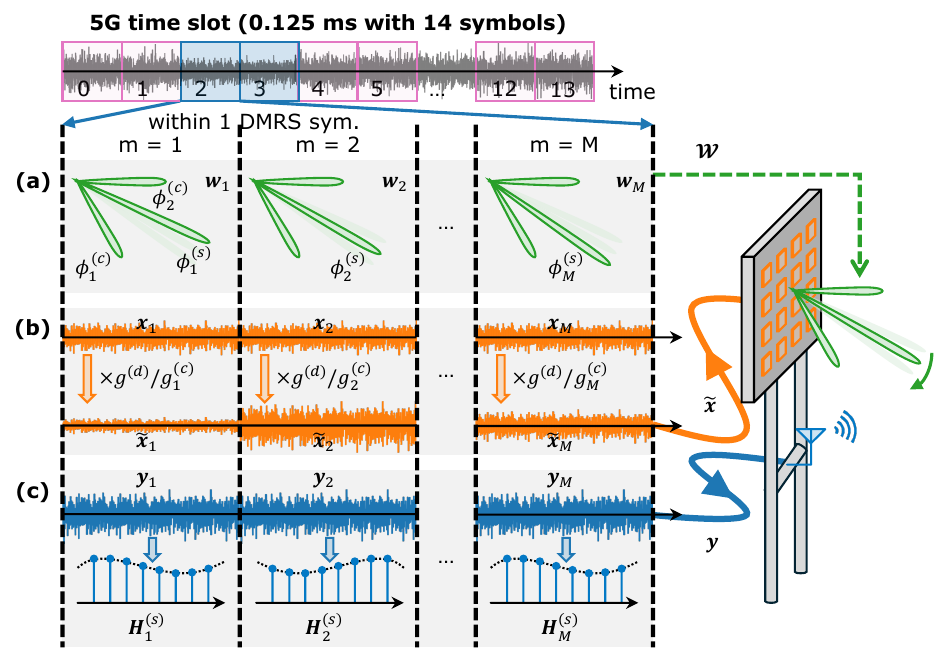}
    \vspace{-3mm}
    \caption{The system overview of {\namebf} with three modules: (a) ISAC beamforming codebook design, (b) DMRS waveform pre-distortion, and (c) sensing CSI estimation.}
    \vspace{-3mm}
    \label{fig:system-overview}
\end{figure}
%% figure ends

We consider sweeping across $\angleSenseNum$ sensing angles within a single DMRS symbol (Fig.~\ref{fig:introduction-overview}), which contains $\fftSize$ I/Q samples excluding the CP.
Therefore, each beamformer lasts for a duration of $\fftSizeLow = \left(\frac{\fftSize}{\angleSenseNum}\right)$ I/Q samples, corresponding to approximately {0.24}\thinspace{\usec} with $\fftSize = {1,024}$ and $\angleSenseNum = {34}$ at a baseband sampling rate of {122.88}\thinspace{MHz} used in our experiments (\S\ref{sec: evaluation}).

{\name} consists of three modules operating at the BS to enable ISAC via sub-symbol beam switching (Fig.~\ref{fig:system-overview}).
The first \emph{ISAC Beamforming Codebook Design} module optimizes the beamformers to be switched during each DMRS symbol to obtain sensing information.
Specifically, each beamformer keeps the beams toward all $\userNum$ users located at $\angleUserAz{\userIdx}$ (to maintain a good communication SNR, $\snrComm{\userIdx}$), while creating an additional sensing beam toward a desired direction $\angleSenseAz$ (to achieve a large sensing SNR, $\snrSense$), that can be rapidly swept even during one DMRS symbol.
The second \emph{DMRS Waveform Pre-Distortion} module adjusts the transmitting gain on the DMRS symbol per switched beamformer to compensate for the beamforming gain mismatch between the data symbol.
In such a way, the users can still rely on the estimated communication CSI for data demodulation.
The third \emph{Sensing CSI Estimation} module estimates the round-trip sub-symbol sensing CSI per beamformer, i.e., per desired sensing angle.

%% figure begins
\begin{figure}[!t]
    \centering
    % \vspace{-3mm}
    \includegraphics[width=0.98\columnwidth]{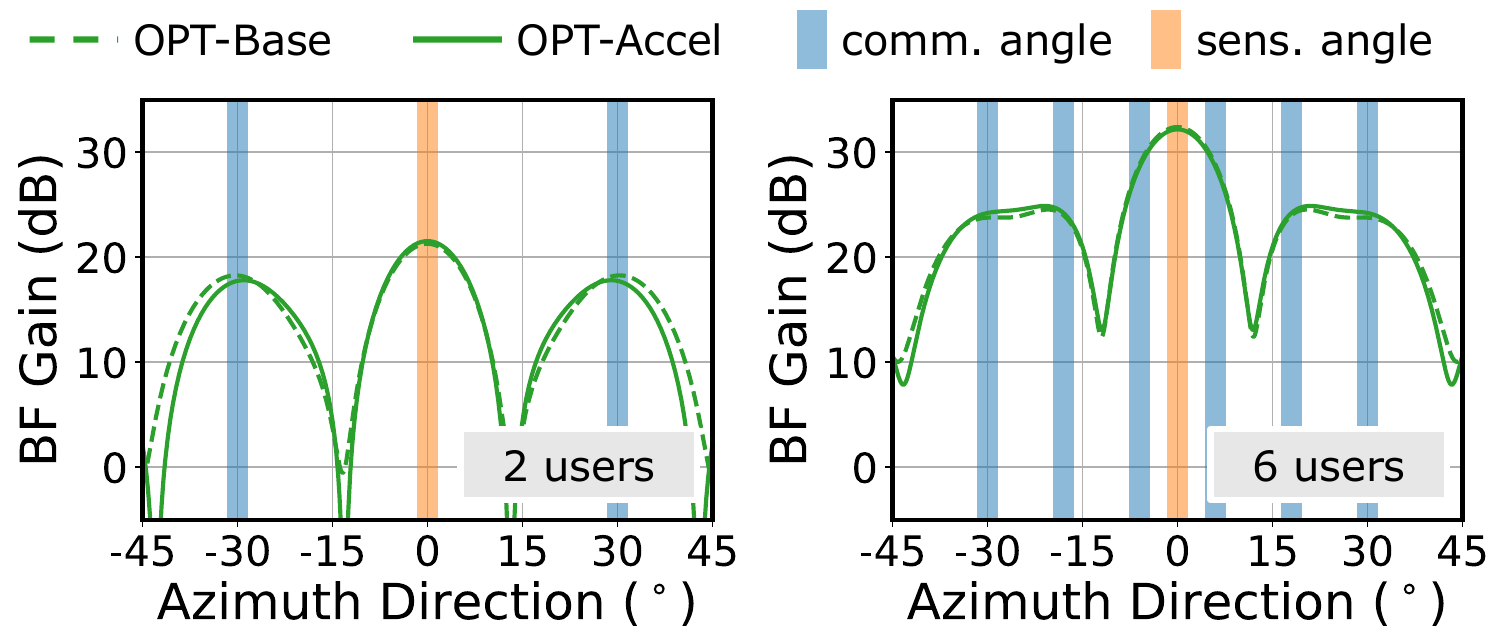}
    \vspace{-3mm}
    \caption{The simulated beamforming optimization comparison between {\BeamOptimOrig} and {\BeamOptim} with two and six users allocated within $\pm 30^{\circ}$ and a sensing beam towards $0^{\circ}$.}
    \vspace{-3mm}
    \label{fig:optimization-compare}
\end{figure}
%% figure ends

\subsection{ISAC Beamforming Codebook Design}
\label{ssec: system-design-beamforming-optimization}

The first module efficiently optimizes and maintains an ISAC beamforming codebook $\bfWeightVecSet = \left\{\bfWeightVec_{\angleSenseIdx}\right\}$, which is employed by the BS to switch across $\angleSenseNum$ angles, $\angleSenseSet = \{ \angleSenseAz_{\angleSenseIdx} \}$, during one DMRS symbol (see Fig.~\ref{fig:system-overview}).
Here, $\bfWeight_{\angleSenseIdx}$ denotes the beamformer applied to the $\angleSenseIdx$-th sub-symbol corresponding to sensing in the direction of $\angleSenseAz_{\angleSenseIdx}$.
The design of $\bfWeightVecSet$ is guided by two key objectives:
(\emph{i}) to generate an additional beam at the desired sensing angle while preserving the beamforming gain toward all the $\userNum$, and
(\emph{ii}) to ensure that the optimization for $\angleSenseNum$ beamformers adapts and converges rapidly enough to keep up with user mobility.

\myparatight{Baseline ISAC beamforming.}
We first formulate a baseline beamforming optimization, {\BeamOptimOrig}, to simultaneously generate one sensing beam alongside $\userNum$ beams toward the $\userNum$ users.
{\BeamOptimOrig} simultaneously maximizes the \emph{communication SNR}, $\snrComm{\userIdx}$ in {\eqref{eq: communication-snr}}, across all users, and \emph{sensing SNR}, $\snrSense$ in {\eqref{eq: sensing-snr}}, with a hyperparameter, $\alpha$, that tunes the balance between the sensing and communication SNR:
\begin{tcolorbox}[boxsep=0mm,left=1mm,right=1mm,top=-1mm,bottom=1mm,
 toptitle=1mm,bottomtitle=1mm,lefttitle=2mm,
 title={{\BeamOptimOrig} to steer the sensing beam towards $\angleSenseAz$},
 boxrule=1pt,sharp corners=all]
\begin{align}
    \bfWeightOpt(\angleSenseAz) &= \arg \max_{\bfWeightVec}: \alpha \cdot \snrSense(\bfWeightVec) + \frac{1}{\userNum} \cdot \sum_{\userIdx=1}^{\userNum} \snrComm{\userIdx}(\bfWeightVec) \\
    \text{s.t.:}~
    & \myAbs{\bfWeight_{\antIdx}} \in [0, 1], ~\forall \antIdx = 1, \dots, \antNum,
    \\
    & \snrSense = \snrBaseSense \cdot \myAbs{\myTrans{\steerVec}(\angleSenseAz) \cdot \bfWeightVec}^2,
    \\
    & \snrComm{\userIdx} = \snrBaseComm{\userIdx} \cdot \myAbs{\myTrans{\steerVec}(\angleUserAz{\userIdx}) \cdot \bfWeightVec}^2, ~\forall \userIdx = 1, \dots, \userNum.
\end{align}
\end{tcolorbox}

\myparatight{Accelerated beamforming codebook construction.}
Directly obtaining $\bfWeightVecSet$ by solving {\BeamOptimOrig} $\angleSenseNum$ times--one for each sensing angle--can be computationally expensive, especially due to its nonlinear constraints.
To address this, we consider an alternative optimization that accelerates the beamforming codebook generation while remaining adaptive to user mobility.
First, instead of maximizing the aggregate user SNR, we re-design the optimization objective to maximize the minimum communication SNR across all users, defined as $\snrCommMin = \min_{\userIdx} \{ \snrComm{\userIdx} \}$. 
This ensures user fairness by effectively mitigating the drawback of {\BeamOptimOrig}, where some users may suffer from poor SNR, thereby hindering reliable communication CSI estimation from the received DMRS symbols.
Second, we introduce a new approach to simultaneously generate the sensing and communication beams.
Specifically, the beamforming weights are initialized using conjugate beamforming based on the target sensing angle $\angleSenseAz$, given by $\bfWeightVec^{\textrm{conj}} = \myConj{\steerVec}(\angleSenseAz)$.
We then perturb $\bfWeightVec^{\textrm{conj}}$ under the constraint that the Euclidean distance of the per-element perturbation remains below a threshold, $\perturb$.
This threshold governs the trade-off between sensing and communication:
a smaller $\perturb$ biases the beamforming weights to preserve the sensing SNR, while a large $\perturb$ allows greater flexibility to redistribute communication SNR across the users.
This approach, {\BeamOptim}, not only balances sensing and communication objectives but also reduces the number of iterations required for convergence, especially with small values of $\perturb$.
\begin{tcolorbox}[boxsep=0mm,left=1mm,right=1mm,top=-1mm,bottom=1mm,
 toptitle=1mm,bottomtitle=1mm,lefttitle=2mm,
 title={{\BeamOptim} to steer the sensing beam towards $\angleSenseAz$},
 boxrule=1pt,sharp corners=all]
\begin{align}
    \bfWeightOpt &= \arg \max_{\bfWeightVec}: \snrCommMin \\
    \text{s.t.:}~
    & \snrComm{\userIdx} = \snrBaseComm{\userIdx} \cdot \myAbs{\myTrans{\steerVec}(\angleUserAz{\userIdx}) \cdot \bfWeightVec}^2 \geq \snrCommMin, ~\forall \userIdx = 1, \dots, \userNum,
    \label{eq:beam-optim-keep-signal} \\
    & \myAbs{\bfWeight_{\antIdx}} \in [0, 1], ~\forall \antIdx = 1, \dots, \antNum,
    \label{eq:beam-optim-amp-limit} \\
    & \myAbs{\bfWeight_{\antIdx} - \bfWeight_{\antIdx}^{\mathrm{conj}}} \leq \perturb, ~\forall \antIdx = 1, \dots, \antNum,
    \label{eq:beam-optim-perturb}\\
    & \bfWeightVec^{\mathrm{conj}} = \myConj{\steerVec} \left(\angleSenseAz \right).
    \label{eq:beam-optim-init}
\end{align}
\end{tcolorbox}

\noindent
Fig.~\ref{fig:optimization-compare} shows simulated beam patterns obtained by {\BeamOptimOrig} and {\BeamOptim} with a fixed sensing angle at $0^{\circ}$, while users are evenly distributed within $\pm 30^{\circ}$.
It can be seen that the resulting beam patterns from the two optimizations are very similar: for each target beamforming and sensing angle, the beamforming gain difference is always less than {1}\thinspace{dB}.

\myparatight{Online beamformer codebook update.}
Next, we design an algorithm for updating the codebook in an online fashion with the presence of mobile users, as summarized in Algorithm~\ref{algo:optimize-beam}.
While {\BeamOptim} must be solved $\angleSenseNum$ times during initialization, it is designed to be more suitable for adapting to user mobility.
As described in Algorithm~\ref{algo:optimize-beam}, when the $\userIdx$-th user moves and its angle changes from $\angleUserAz{\userIdx}$ to $\angleUserNew{\userIdx}$, the set of beamforming weights $\bfWeightVecSet$ across all sensing angles must be updated accordingly to maintain a good communication SNR to the user.
Fortunately, under {\BeamOptim}, most of the beamformers in the previous codebook can be directly reused in the updated codebook.
For each existing beamformer $\bfWeightOpt_{\angleSenseIdx}$, we evaluate the updated user's SNR, $\snrCommNew{\userIdx}$,
by plugging the new user angle $\angleUserNew{\userIdx}$.
If the updated SNR $\snrCommNew{\userIdx}$ does not degrade the optimization objective in {\eqref{eq:beam-optim-keep-signal}}, then $\bfWeightOpt_{\angleSenseIdx}$ remains a valid optimal solution to {\BeamOptim}.
This argument can be proved by contradiction: if there exist such beamforming weights producing a larger objective $\snrCommMin_{\angleSenseIdx}$ in {\BeamOptim}, i.e., $\snrCommNew{\userIdx} \geq \snrCommMin_{\angleSenseIdx}$, it must satisfy all the constraints in the old {\BeamOptim} since the only updated $\snrCommNew{\userIdx}$ does not influence $\snrCommMin_{\angleSenseIdx}$, which conflicts with the optimal solution of the old {\BeamOptim}.
Otherwise, for $\snrCommNew{\userIdx} < \snrCommMin_{\angleSenseIdx}$, we need to fine-tune the beamforming weights to incorporate the user angle change. The beamforming weights are clipped into a small range around the conjugate beamforming $\bfWeightVec^{\textrm{conj}}$, which means the convergence can still be quickly reached.

\begin{algorithm}[!t]
\small
\KwIn{Old $\bfWeightVecSet$ and $\snrCommMinSet$, the new user angle(s) at $\angleUserNew{\userIdx}$}
\KwOut{New $\bfWeightVecSet$ and $\snrCommMinSet$}
\SetKwBlock{Begin}{function}{end function}

\For{$\angleSenseAz_{\angleSenseIdx} \in \angleSenseSet$}
{
    Load the old $\bfWeightOpt_{\angleSenseIdx}$ from $\bfWeightVecSet$; 
    load the old $\snrCommMin_{\angleSenseIdx}$ from $\snrCommMinSet$ \\
    Calculate $\snrCommNew{\userIdx}$ with old $\bfWeightOpt_{\angleSenseIdx}$ and new $\angleUserNew{\userIdx}$ based on {\eqref{eq: communication-snr}} \\
    \If{$\myAbs{\min_{\userIdx} \snrCommNew{\userIdx} - \snrCommMin_{\angleSenseIdx}} \leq 10^{-2}$}
    {
        No need to update $\bfWeightOpt_{\angleSenseIdx}$ \\
    }
    \Else{
        Initialize $\bfWeightOpt_{\angleSenseIdx}$ by the old $\bfWeightOpt_{\angleSenseIdx}$ \\
        \While{$\myAbs{\nabla \snrCommMin_{\angleSenseIdx}} \geq 10^{-2}$}
        {
            Iterate the new $\bfWeightOpt_{\angleSenseIdx}$,
            update the new $\snrCommMin_{\angleSenseIdx}$ \\
        }
        Save the new $\bfWeightOpt_{\angleSenseIdx}$ to $\bfWeightVecSet$; 
        save the new $\snrCommMin_{\angleSenseIdx}$ to $\snrCommMinSet$ \\
    }
}
\Return{$\bfWeightVecSet, \snrCommMinSet$}

\caption{Online beamformer update.}
\label{algo:optimize-beam}
\end{algorithm}

\subsection{DMRS Waveform Pre-Distortion}
\label{ssec: system-design-communication-csi}

The second module pre-distorts the DMRS waveform for each sub-symbol with a different beamformer, which ensures a consistent beamforming gain toward the user across the entire DMRS symbol.
In this way, the user can follow the conventional procedure specified by the 5G protocol for estimating the communication CSI.

Let $\bfGainData{\userIdx}$ denote the beamforming gain of the data symbols for the $\userIdx$-th user.
For the $\angleSenseIdx$-th sub-symbol within the DMRS symbol, the beamforming gain towards the $\userIdx$-th user can be derived by $\bfGainComm{\angleSenseIdx, \userIdx} = \myAbs{\myTrans{\steerVec}(\angleUserAz{\userIdx}) \cdot \bfWeightVec_{\angleSenseIdx}}^2$.
The desired pre-distortion factor for the $\angleSenseIdx$-th sub-symbol is given by the beamforming gain ratio, $\bfGainData{\userIdx} / \bfGainComm{\angleSenseIdx, \userIdx}$, which is user-dependent. On the other hand, only a single factor can be applied to the $\angleSenseIdx$-th sub-symbol.
Interestingly, we observe that this beamforming gain shows minimal variation across users.
Therefore, we take the average beamforming gain over the $\userNum$ users as $\bfGainDataMean$ for data symbols and $\bfGainCommMean{\angleSenseIdx}$ for the $\angleSenseIdx$-th sub-symbol, and then derive the beamforming gain ratio as the pre-distortion factor.
For the $\angleSenseIdx$-th sub-symbol, the pre-distortion process from $\waveTxLowVec{\angleSenseIdx}$ to $\waveTxVecCal_{\angleSenseIdx}$ can be formulated as (see also Fig.~\ref{fig:system-overview}):
\begin{align}
    \textstyle \waveTxVecCal_{\angleSenseIdx}
    & = \sqrt{\frac{\bfGainDataMean}{\bfGainCommMean{\angleSenseIdx}}} \cdot \waveTxLowVec{\angleSenseIdx} = \sqrt{\frac{\sum_{\userIdx} \bfGainData{\userIdx}}{\sum_{\userIdx} \bfGainComm{\angleSenseIdx, \userIdx}}} \cdot \waveTxLowVec{\angleSenseIdx},\ \forall m = 1, \dots, \angleSenseNum.
\end{align}
where the square root is to convert the power ratio to the amplitude ratio.
In this way, each user can perform conventional CSI estimation as discussed in \S\ref{ssec: preliminaries-dmrs-csi} for data demodulation.

\subsection{Sensing CSI Estimation}
\label{ssec: system-design-sensing-csi}

\begin{figure}[!t]
    \begin{minipage}{0.49\columnwidth}
        \centering
        \includegraphics[width=1\linewidth]{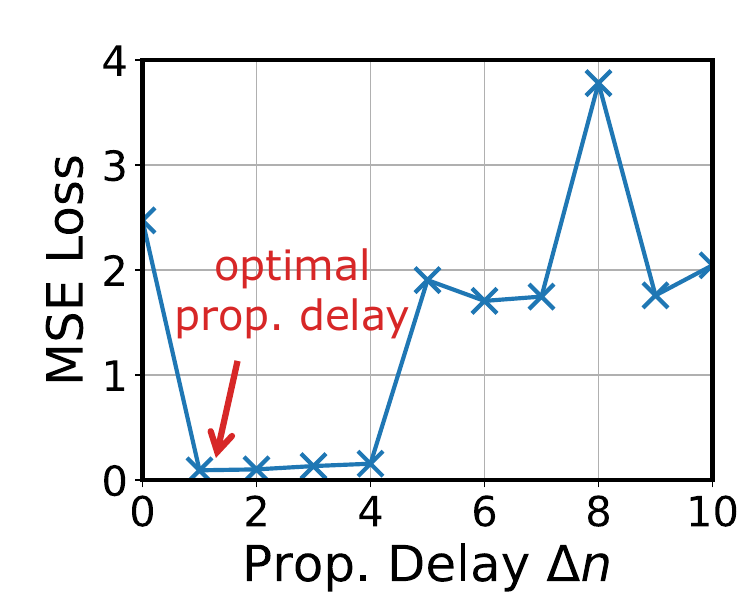}
        \vspace{-3mm}
        \caption{The optimal propagation delay given by minimizing the MSE loss.}
        \vspace{-3mm}
        \label{fig: CSI-example}
    \end{minipage}
    \hfill
    \begin{minipage}{0.49\columnwidth}
        \centering
        \includegraphics[width=1\linewidth]{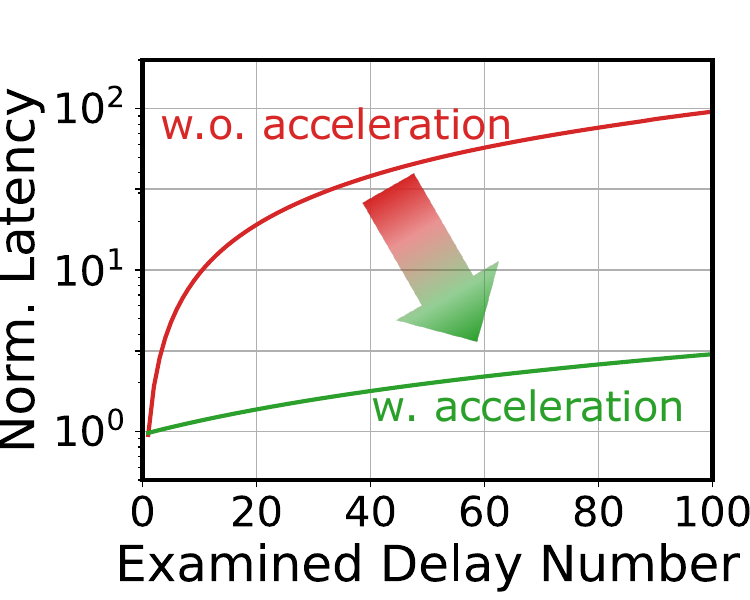}
        \vspace{-3mm}
        \caption{The latency reduction by the queue-based propagation delay estimation.}
        \vspace{-3mm}
        \label{fig: fft-acceleration}
    \end{minipage}
\end{figure}

On the BS side, we extract sub-symbol sensing CSI per beamformer with respect to each sensing angle, $\angleSenseAz_{\angleSenseIdx}$.

\myparatight{Received waveform propagation delay optimization.}
We assume the synchronized beamforming steering and the data streams on the BS, which means we can precisely locate the waveform transmitted within a given sub-symbol beamformer.
The key challenge of the sensing CSI estimation is to estimate the unknown propagation delay for aligning the received and transmitted waveforms.
Specifically, the transmitted waveform under the $\angleSenseIdx$-th beamformer, $\waveTxLowVec{\angleSenseIdx}$, is an $\fftSizeLow$-point sub-vector of $\waveTxVec$ indexed from $\left[ (\angleSenseIdx-1)\fftSizeLow+1, \dots, \angleSenseIdx\fftSizeLow \right]$; assuming a propagation delay of $\waveDelay$ I/Q samples, its corresponding received waveform $\waveRxLowVec{\angleSenseIdx}$ is an $\fftSizeLow$-point sub-vector of $\waveRxVec$ indexed from $\left[(\angleSenseIdx-1)\fftSizeLow+1+\waveDelay, \dots, (\angleSenseIdx\fftSizeLow+\waveDelay)\right]$.
Under this assumed propagation delay $\waveDelay$, the corresponding sensing CSI $\CSILowVec{\angleSenseIdx}(\waveDelay) = [\CSILow{\angleSenseIdx}(\waveDelay)[\specIdx]] \in \mathbb{C}^{\fftSizeLow}$ for the $\angleSenseIdx$-th direction is given by
\begin{align}
    \CSILow{\angleSenseIdx}(\waveDelay)[\specIdx] &= \frac{\specRxLow{\angleSenseIdx}(\waveDelay)[\specIdx]}{\specTxLowCal{\angleSenseIdx}[\specIdx]} = \sqrt{\frac{\bfGainCommMean{\angleSenseIdx}}{\bfGainDataMean}} \cdot \frac{\fft(\waveRxLowVec{\angleSenseIdx}(\waveDelay))[\specIdx]}{\fft(\waveTxLowVec{\angleSenseIdx})[\specIdx]},
    \label{eq: sensing-csi-estimation}
\end{align}
where we replace $\waveTxVec_{\angleSenseIdx}$ (or $\specTxLowVec{\angleSenseIdx}$) by the pre-compensated waveform $\waveTxVecCal_{\angleSenseIdx}$ (or $\specTxLowVecCal{\angleSenseIdx}$) to include the beamforming gain ratio.
To estimate the propagation delay $\waveDelayOpt_{\angleSenseIdx}$, we examine each sensing CSI candidate $\CSILowVec{\angleSenseIdx}(\waveDelay)$ by sweeping different values of $\waveDelay$.
The likelihood of the actual propagation delay being $\waveDelayOpt_{\angleSenseIdx}$ is given by the assumption of {\eqref{eq: channel-model}}, where only the path within the dedicated sensing angle $\angleSenseAz_{\angleSenseIdx}$ dominates the channel.
Based on this assumption, mathematically, the phases of the estimated sensing CSI over subcarriers should follow the \emph{sawtooth} shape as illustrated in \S\ref{ssec: preliminaries-dmrs-csi}, when the correct propagation delay $\waveDelay = \pathDelay_{\textrm{max}} \cdot \sampRate$ is plugged so that $\waveTxLowVec{\angleSenseIdx}$ and $\waveRxLowVec{\angleSenseIdx}(\waveDelay)$ are exactly aligned.
The propagation delay estimation of $\waveDelayOpt_{\angleSenseIdx}$ can be formulated as
\begin{tcolorbox}[boxsep=0mm,left=1mm,right=1mm,top=-1mm,bottom=1mm,
 toptitle=1mm,bottomtitle=1mm,lefttitle=2mm,
 title={{\DelayOptim} to optimize the propagation delay $\waveDelay$},
 boxrule=1pt,sharp corners=all]
\begin{align}
    \waveDelayOpt_{\angleSenseIdx} &= \arg \min_{\waveDelay, \phaseSlope, \phaseBias} \myNorm{ \myAbs{\specTxLowVec{\angleSenseIdx}} \odot \left(\myAng{\CSILowVec{\angleSenseIdx}}(\waveDelay) - \myAng{\hat{\CSIVec}(\phaseSlope, \phaseBias)} \right) }^2 \\
    \text{s.t.:}~
    &\CSILow{\angleSenseIdx}(\waveDelay)[\specIdx] = \sqrt{\frac{\bfGainCommMean{\angleSenseIdx}}{\bfGainDataMean}} \cdot \frac{\fft(\waveRxLowVec{\angleSenseIdx}(\waveDelay))[\specIdx]}{\fft(\waveTxLowVec{\angleSenseIdx})[\specIdx]}, \nonumber \\
    &\quad \quad \quad \forall \specIdx=1, \dots, \fftSizeLow, \\
    &\myAng{\hat{\CSI}}[\specIdx] = \phaseSlope \cdot \specIdx + \phaseBias, ~\forall \specIdx=1, \dots, \fftSizeLow, \\
    &\waveDelay = 0, 1, \dots, \waveDelayNum-1.
\end{align}
\end{tcolorbox}

\noindent
Here, `$\odot$' denotes the element-wise multiplication, and $\myNorm{\cdot}$ denotes the $L$-2 norm of an $\fftSizeLow$-point vector.

{\DelayOptim} is solved using a two-stage process.
First, for a given delay $\waveDelay$, we compute the corresponding CSI vector $\CSILowVec{\angleSenseIdx}(\waveDelay)$ from {\eqref{eq: sensing-csi-estimation}}.
This CSI is then fitted to a linear phase profile, denoted as $\hat{\CSIVec} = [\hat{\CSI}[\specIdx]] \in \mathbb{C}^{\fftSizeLow}$, parametrized by $\phaseSlope$ and $\phaseBias$ (or appearing as sawtooth pattern when wrapped into $[0, 2\pi)$) .
We element-wise weight the importance of each subcarrier by the transmitted amplitude, $\myAbs{\CSILowVec{\angleSenseIdx}}$, therefore excluding unused or low-power subcarriers.
Second, we enumerate candidate delay values $\waveDelay=0, 1, \dots, (\waveDelayNum-1)$,
and, for each $\waveDelay$, evaluate the goodness of fit using mean squared error (MSE) between the fitted phase $\myAng{\hat{\CSIVec}(\phaseSlope, \phaseBias)}$ and that of the CSI $\myAng{\CSILowVec{\angleSenseIdx}}(\waveDelay)$.
The propagation delay $\waveDelay$ that minimizes this MSE is selected as the optimal estimate, denoted by $\waveDelayOpt$.
An experimental example of the MSE loss profile over varying propagation delay $\waveDelay$ is shown in Fig.~\ref{fig: CSI-example}.
The resulting sensing CSI for the $\angleSenseIdx$-th sensing angle is then given by $\CSILowVecOpt{\angleSenseIdx} = \CSILowVec{\angleSenseIdx}(\waveDelayOpt_{\angleSenseIdx})$.
This process is repeated for all $\angleSenseNum$ sensing angles to obtain $\angleSenseNum$ sensing CSI estimates during one DMRS symbol.

\myparatight{Queue-based propagation delay estimation acceleration.}
We further develop a queue-based algorithm to accelerate {\DelayOptim}.
First, the per-sub-symbol beamformer $\specTxLowVec{\angleSenseIdx}$ of the transmitted waveform can be preprocessed offline.
In the online phase, the linear form of the optimization ensures the parameters $\phaseSlope$ and $\phaseBias$ can be solved in an analytical way with ignorable computation cost.
Therefore, the computation cost is bottlenecked by the $\fftSizeLow$-point FFT that converts $\waveRxLowVec{\angleSenseIdx}(\waveDelay)$ to $\specRxLowVec{\angleSenseIdx}(\waveDelay)$ per given $\waveDelay$.
In our queue-based acceleration, we start with the first propagation delay $\waveDelay=0$, where the $\fftSizeLow$ I/Q samples as of $\waveRxLowVec{\angleSenseIdx}(0)$ are enqueued. Its corresponding $\specRxLowVec{\angleSenseIdx}(0)$ can be calculated by an $\fftSizeLow$-point FFT over the queue at the computation complexity of $\complexity{\fftSizeLow \log \fftSizeLow}$.
Then, for the successive $\waveDelay$, we notice that, its waveform $\waveRxLowVec{\angleSenseIdx}(\waveDelay)$ is offset to the last waveform $\waveRxLowVec{\angleSenseIdx}(\waveDelay-1)$ by one I/Q sample. Hence, the queue can be updated by dequeuing the first I/Q sample in $\waveRxLowVec{\angleSenseIdx}(\waveDelay-1)$, denoted as $\waveRx_{\textrm{out}}$, and enqueuing the last I/Q sample in $\waveRxLowVec{\angleSenseIdx}(\waveDelay)$, denoted as $\waveRx_{\textrm{in}}$.
Comparing the FFT formula of the two waveforms, such a dequeue and enqueue operation is reflected on the conversion from $\specRxLowVec{\angleSenseIdx}(\waveDelay-1)$ to $\specRxLowVec{\angleSenseIdx}(\waveDelay)$ as
\begin{align}
    \specRxLow{\angleSenseIdx}(\waveDelay)[\specIdx] &= \left( \specRxLow{\angleSenseIdx}(\waveDelay-1)[\specIdx] + \waveRx_{\textrm{in}} - \waveRx_{\textrm{out}} \right) \cdot \eu^{\iu 2\pi \frac{\specIdx}{\fftSizeLow}}.
\end{align}
In this way, the computation complexity of acquiring the new $\specRxLowVec{\angleSenseIdx}(\waveDelay)$ is reduced from $\complexity{\fftSizeLow \log \fftSizeLow}$ to $\complexity{\fftSizeLow}$.

\myparatight{Computation complexity analysis.}
For each {\DelayOptim}, it takes $\complexity{\fftSizeLow \log \fftSizeLow}$ for FFT to calculate $\specRxLow{\angleSenseIdx}(0)$ with $\waveDelay=0$, and it takes $\complexity{\fftSizeLow}$ to iteratively update the following $\specRxLow{\angleSenseIdx}(\waveDelay)$ from $\specRxLow{\angleSenseIdx}(\waveDelay-1)$ for $\waveDelay>0$.
Consider a total of $\waveDelayNum$ trials for the $\waveDelay$, the overall computation complexity is given by $\complexity{\fftSizeLow \log \fftSizeLow + \fftSizeLow \cdot \waveDelayNum}$.
For the whole DMRS symbol with $\angleSenseNum$ beamformers, the total computation complexity becomes $\complexity{\fftSize \log \fftSizeLow + \fftSize \cdot \waveDelayNum}$.
In comparison, the conventional CSI estimation over the whole $\fftSize$ I/Q samples of the DMRS symbol is with the computation complexity of $\complexity{\fftSize \log \fftSize}$.
Hereby, we empirically choose $\waveDelayNum=\log \fftSize$ so that our sensing CSI estimation has the same computation complexity as the conventional way.
Fig.~\ref{fig: fft-acceleration} shows the running latency over different $\waveDelayNum$ between {\name}'s sensing CSI estimation with/without the acceleration and the conventional CSI estimation.

\section{Implementation}
\label{sec:implementation}

\subsection{PAWR COSMOS Testbed}

%% figure begins
\begin{figure}[!t]
    \centering
    % \vspace{-3mm}
    \includegraphics[width=0.98\columnwidth]{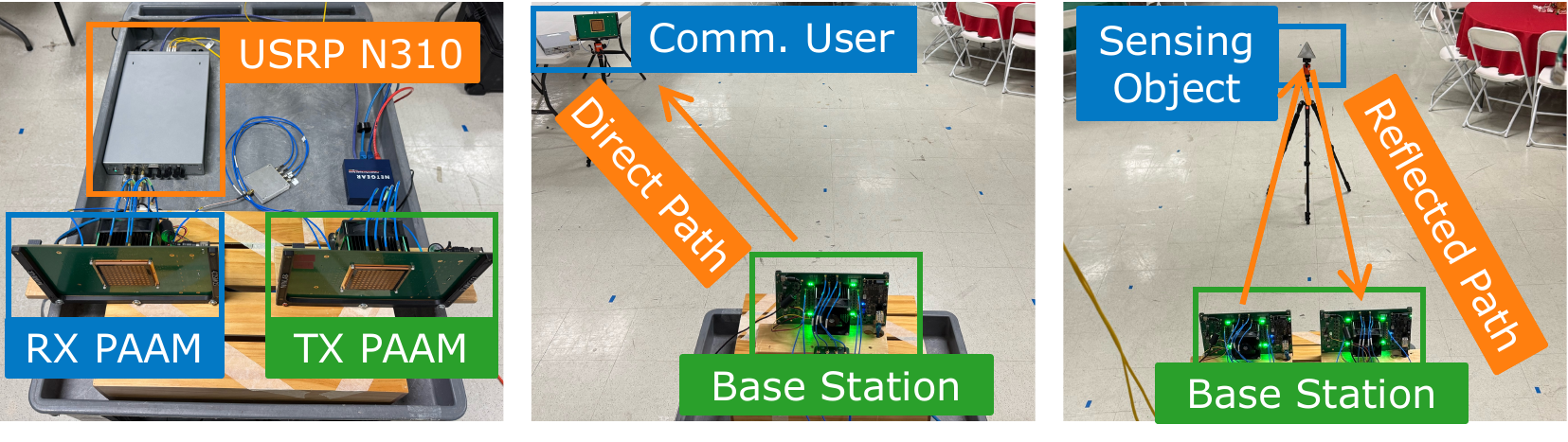}
    \vspace{-3mm}
    \caption{Two IBM {28}\thinspace{GHz} PAAM boards connected to USRP N310 SDRs, deployed in the PAWR COSMOS testbed.}
    \vspace{-3mm}
    \label{fig:experiment-setup}
\end{figure}
%% figure ends

We implement and evaluate {\name} using the IBM {28}\thinspace{GHz} PAAMs~\cite{gu2021development, chen2023open} and USRP N310s SDRs~\cite{usrp_n310} deployed in the PAWR COSMOS testbed~\cite{raychaudhuri2020challenge}.
Fig.~\ref{fig:experiment-setup} shows the experimental setup.
The IBM {28}\thinspace{GHz} PAAM is an 8$\times$8 antenna array front-end hardware with an array of subarrays (ASA) architecture.
It comprises four subarrays, each connected to an independent TX and RX chain at {3}\thinspace{GHz} intermediate frequency (IF). 
Each subarray consists of a 4$\times$4 antenna array with element spacing at approximately half-wavelength.
The typical field of view (FoV) of the antenna spans from {-60}$^{\circ}$ to {+60}$^{\circ}$.
Each antenna element supports programmable beamforming weights with 5-bit amplitude resolution ($\myAbs{\bfWeight_{\antIdx}}$) and a phase resolution of {4.87}$^{\circ}$ ($\myAng{\bfWeight_{\antIdx}}$). 
Most relevant to {\name}, the beamforming weights can be reconfigured as quickly as every {0.24}\thinspace{\usec}, enabling sub-symbol beam switching.
The USRP N310~\cite{usrp_n310} is a high-performance SDR that provides IF signals from/to PAAM at {3}\thinspace{GHz}.
It is equipped with a 14-bit DAC and a 16-bit ADC, and operates at a baseband sampling rate of {122.88}\thinspace{MHz}.
The USRP N310 is controlled via the Python-based GNU Radio interface over a 10\thinspace{Gbps} SFP+ link.

\subsection{Beam Pattern Measurements}

To evaluate {\name}'s beamforming optimization algorithm {\BeamOptim}, we measure the transmit beam pattern in the horizontal plane in an anechoic chamber, demonstrating simultaneously beams for communication and sensing.
For the TX PAAM, each subarray is driven by a signal generator that provides a {3}\thinspace{GHz} IF tone for up-conversion to {28}\thinspace{GHz}. 
A horn antenna is placed {1.5}\thinspace{m} broadside to the transmitter, and is connected to a spectrum analyzer to capture the over-the-air mmWave signals.
The TX PAAM is mounted on a programmable motor that rotates across the azimuth angles in {0.1}$^{\circ}$ increments, so that the receiver can record the beam pattern.
Due to the mmWave absorbers filled in this anechoic chamber, multi-path reflections are suppressed, ensuring that the beamforming gain can be accurately obtained from the measured signal power.

\subsection{System-Level Implementation}
\label{ssec: implementation-system-implementation}

We evaluate the performance of {\name}, including its key modules, in an open environment, as shown in Fig.~\ref{fig:experiment-setup}.

\myparatight{5G PDSCH waveform generation.}
We establish a PDSCH data link under 5G FR2 with numerology 3 (NU3)~\cite{MATLAB5GToolbox}, corresponding to a subcarrier spacing of {120}\thinspace{kHz} and slot duration of {0.125}\thinspace{ms}.
Each time slot contains 14 OFDM symbols, and each symbol lasts {8.33}\thinspace{\usec} excluding the CP; the {3/4/11/12}-th symbols are designated as the DMRS symbols, and others are data symbols.
We consider an FFT size of {1,024}, i.e., a sampling rate of {122.88}\thinspace{MHz}, and only the middle {768} subcarriers are occupied, resulting in an occupied bandwidth of {92.16}\thinspace{MHz}.
We randomize the data bits for communication, and vary over {28} modulation and coding schemes (MCSs) following the 5G MCS table~\cite{sharetechnote, chen2005improved} from QPSK-$\big(\frac{120}{1,024}\big)$ ({0.022}\thinspace{Gbps} per data stream) to 256QAM-$\big(\frac{948}{1,024}\big)$ (i.e., {0.683}\thinspace{Gbps} per data stream). The encoding/decoding process is based on the low-density parity-check (LDPC) coding following the 5G standards.
For each MCS, we experiment with the minimum link SNR such that the link block error rate (BLER) is below {10\%}.
The generated waveform is transmitted and received by the pair of PAAM and USRP N310 at {122.88}\thinspace{MHz} sampling rate through an over-the-air {28}\thinspace{GHz} channel.

\myparatight{BS setup.}
On the BS side, we deploy a pair of TX and RX PAAMs, both connected to the same USRP N310, as shown in Fig.~\ref{fig:experiment-setup}(\emph{left}).
On the TX PAAM, we combine the two horizontal subarrays to form a larger subarray with 8 elements in a row, resulting in a HPBW of {12.3}$^{\circ}$ in the azimuth plane.
During the transmission of DMRS symbols, we leverage fast beam switching offered by PAAM, with each beamformer only lasting for {0.24}\thinspace{\usec}~\cite{chen2023open}. 
This allows the switching of up to {34} beamformers within each DMRS symbol.
As for the data symbol, we exploit the state-of-the-art multi-user communication~\cite{gao2024mambas} to generate the fixed beamformers, and are demodulated by the communication CSI estimated from the pre-distortion DMRS waveform, as described in \S\ref{ssec: system-design-communication-csi}.
The RX PAAM performs fixed conjugate beamforming using a 4$\times$4 subarray toward the broadside direction.

\myparatight{Multi-user communication setup.}
As shown in Fig.~\ref{fig:experiment-setup}(\emph{middle}), we establish a wireless link at a distance of {5}\thinspace{m} to evaluate {\name}'s communication performance.
We consider a single-data-stream user on an additional pair of RX PAAM and USRP N310, where a 4$\times$4 subarray performs fixed conjugate beamforming toward the BS.
Note that the user is not synchronized with the BS in both time and frequency.
Hereby, the CSI is estimated on the users following the conventional 5G CSI estimation (see \S\ref{ssec: preliminaries-dmrs-csi}).
We apply the estimated CSI on the data symbols and report the link performance by error vector magnitude (EVM); after the decoding process, we further report bit error rate (BER) and the maximum achievable MCS based on the block error rate (BLER) as the metrics.

\myparatight{Monostatic sensing setup.}
{\name} employs the monostatic sensing model, where the RX PAAM receives the reflected signals from the surrounding objects for sensing purposes, as shown in Fig.~\ref{fig:experiment-setup}(\emph{right}).
Note that the TX and the RX PAAMs are connected to the same USRP N310 so that they are synchronized.
We keep a distance between the two PAAMs to reduce the self-interference directly from the TX PAAM to the RX PAAM.
As a result, the residual interference-to-noise ratio (INR) is approximately {20}\thinspace{dB}.
This RX PAAM is only active during the DMRS symbols, where the sensing CSI is extracted per switched beamformer.

\myparatight{Sensing CSI features.}
Within each beamformer, we extract three features from the sensing CSI: (\emph{i}) overall received power, which is the sum power over all the subcarriers in the sensing CSI, (\emph{ii}) phase slope, which indicates the propagation delay given the single-path domination assumption, and (\emph{iii}) the linearity loss, indicating how the strongest path dominates the multi-path effect.

\subsection{Communication and Sensing Baselines}
\label{ssec: implementation-baselines}

We implement three communication baselines to benchmark {\name}'s communication and/or sensing performance.

\myparatight{Single-user beamforming (SUBF).}
We first implement the single-user beamforming (SUBF)~\cite{shepard2012argos} widely used in today's 5G networks.
In this baseline, the TX PAAM performs conjugate beamforming with $\bfWeight=\myConj{\steerVec}(\angleUserAz{})$ toward a single user at $\angleUserAz{}$ during both the DMRS and data symbols, which maximizes the received SNR and the data rate at that user.
The RX PAAM at the BS estimates the round-trip CSI from the DMRS symbols for sensing.

\myparatight{Multi-user beamforming (MUBF).}
We also implement {Mambas}~\cite{gao2024mambas} as the multi-user beamforming (MUBF) baseline for communication.
Specifically, the beamformer is optimized with beams toward multiple users with minimal cross-user interference, and is applied to both the data and DMRS symbols.
This communication achieves the state-of-the-art communication sum data rate over multiple users.

\myparatight{Beamformer switching during SSB.}
We also implement a sensing baseline following EyeBeam~\cite{paidimarri2024eye}, which switches beamformers across multiple sensing angles by conjugate beamforming during SSB.
Specifically, each SSB packs {64} beamformers to explore {64} sensing angles, which has an interval of {20}\thinspace{ms}~\cite{sharetechnote}.
The CSI is extracted per switched beamformer for sensing purposes.
As a result, the beamformer switching of {\name} is {340}$\times$ faster than EyeBeam.
\section{Evaluation}
\label{sec: evaluation}

In this section, we evaluate the ISAC beamforming capabilities of {\name} through beam pattern measurements in an anechoic chamber, and its communication and sensing performance in an open environment.
We further conduct runtime experiments in a dynamic setting with mobile user(s) to demonstrate {\name}'s real-time adaptability.

\subsection{ISAC Beamforming}
\label{ssec: evaluation-beamforming}

We first benchmark the beamforming optimization, {\BeamOptim}, by measuring the beamforming pattern in the chamber.

%% figure begins
\begin{figure}%[!t]
    \centering
    \includegraphics[width=0.98\columnwidth]{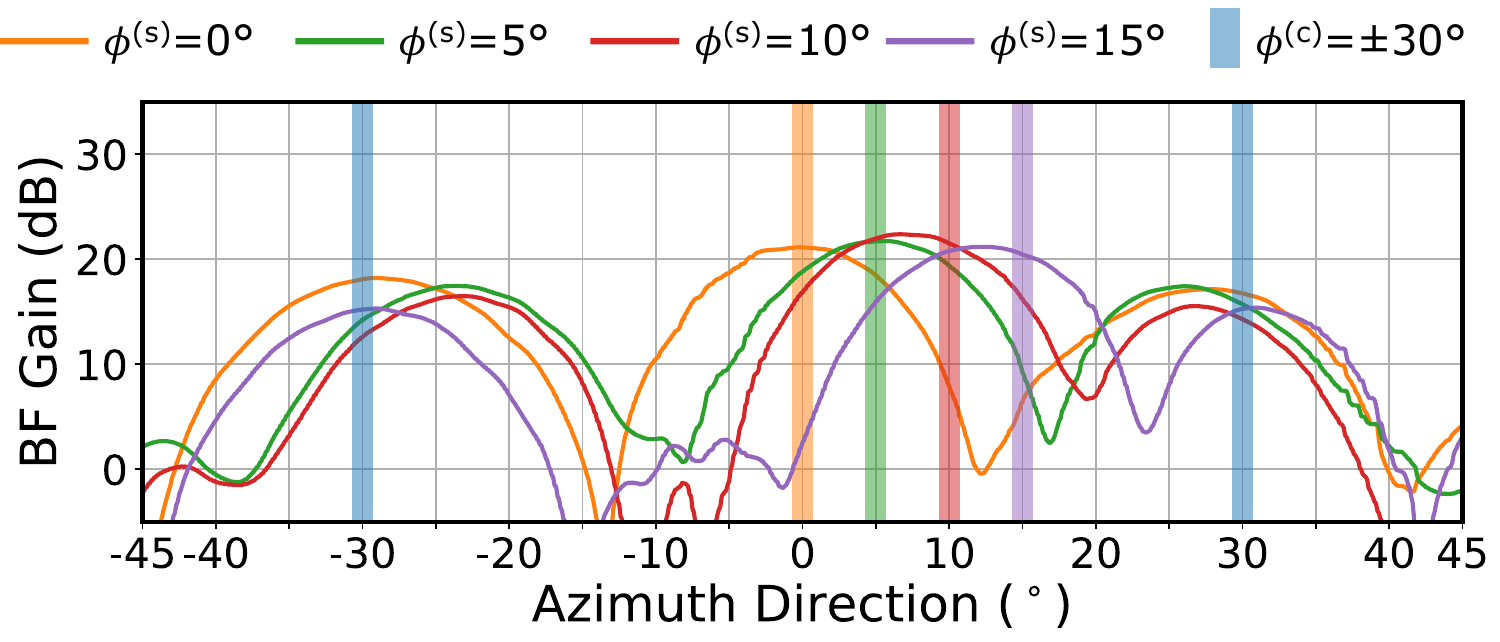}
    \vspace{-3mm}
    \caption{The beamforming pattern with {0/5/10/15}$^{\circ}$ sensing angles, while keeping the communication angle at $\pm 30^{\circ}$.}
    \vspace{-3mm}
    \label{fig:beam-pattern-switching}
\end{figure}
%% figure ends

\myparatight{Beamformers with switching sensing angles.}
We start with the two {1}$\times${8} subarrays serving two users located at $\pm 30^{\circ}$ with an equal baseline SNR of $\snrBaseComm{\userIdx}$.
Then, we optimize the beamformer codebook $\bfWeightVecSet$ with the sensing angle switching across $\angleSenseSet = \{ 0^{\circ}, 5^{\circ}, 10^{\circ}, 15^{\circ}\}$ with $\perturb=0.5$.
Fig.~\ref{fig:beam-pattern-switching} shows the measured beamforming gain over the azimuth angles of the four beamformers.
The measured beamforming gains toward the four target sensing angles are {21.14/21.69/21.56/20.52}\thinspace{dB}, and the corresponding average beamforming gains toward the two users at $\pm 30^{\circ}$ are {17.43/15.01/13.46/15.23}\thinspace{dB}.
These results demonstrate that stable beamforming gains and communication SNRs can be maintained for users, even in the presence of sub-symbol beam switching.

%% figure begins
\begin{figure}%[!t]
    \centering
    \includegraphics[width=0.98\columnwidth]{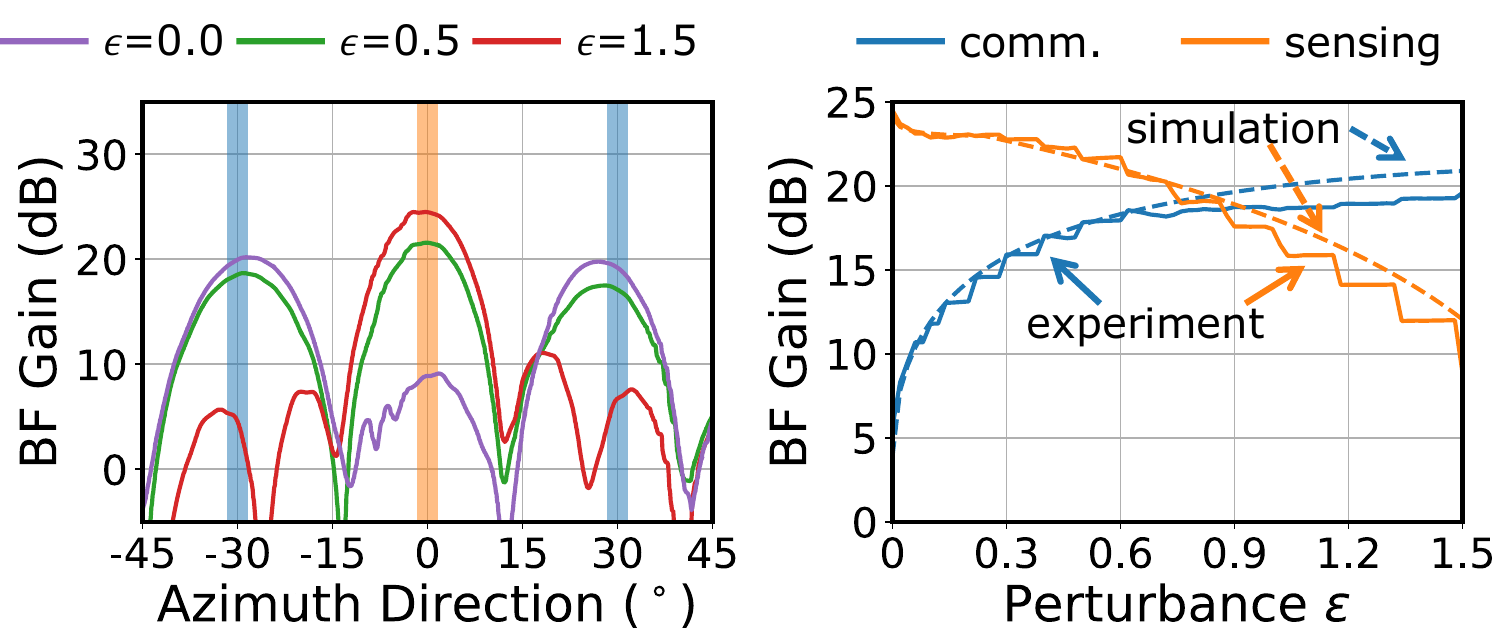}
    \vspace{-3mm}
    \caption{The beamforming patterns with $\perturb=0.5/1.0/1.5$ (left), and the communication and sensing trade-off (right).}
    \vspace{-3mm}
    \label{fig:sensing-communication-tradeoff}
\end{figure}
%% figure ends

\myparatight{Communication and sensing trade-offs.}
{\BeamOptim} provides the hyper-parameter $\perturb$ to balance the beamforming gain between the communication and sensing functionalities.
Under the same setup above with the sensing angle fixed at $\angleSenseAz=0^{\circ}$, Fig.~\ref{fig:sensing-communication-tradeoff}(\emph{left}) shows the detailed beamforming patterns optimized with $\perturb \in \{0, 0.5, 1.5\}$.
Specifically, $\perturb=0$ means the purely conjugate beamforming for the sensing beam at $\angleSenseAz$, whose theoretical beamforming gain toward $\angleSenseAz$ is {24.08}\thinspace{dB}; on the other hand, $\perturb=1.5$ means maximizing the communication beamforming gain for the users, i.e., {20.06/19.37}\thinspace{dB}, therefore sacrificing the sensing performance.
In comparison, we found that applying $\perturb=0.5$ effectively balances both functionalities, i.e., with only {2.92}\thinspace{dB} degradation on the sensing beamforming gain compared to the pure sensing beamformer, and {1.43/2.10}\thinspace{dB} degradation on the communication beamforming gain toward the two users compared to the pure communication beamformer.
Moreover, we show the beamforming gains for sensing and communication with varying $\perturb$ in Fig.~\ref{fig:sensing-communication-tradeoff}(\emph{right}).
Overall, the measurements closely match the simulation results, with the observed step-like pattern arising from beamforming weight quantization.
When setting $\perturb=${0.76}, the beamforming gains for sensing and communication are identical.
In practice, since the base SNR of reflected sensing signals is typically (much) lower, we empirically set $\perturb=0.5$.

%% figure begins
\begin{figure}%[!t]
    \centering
    \includegraphics[width=0.98\columnwidth]{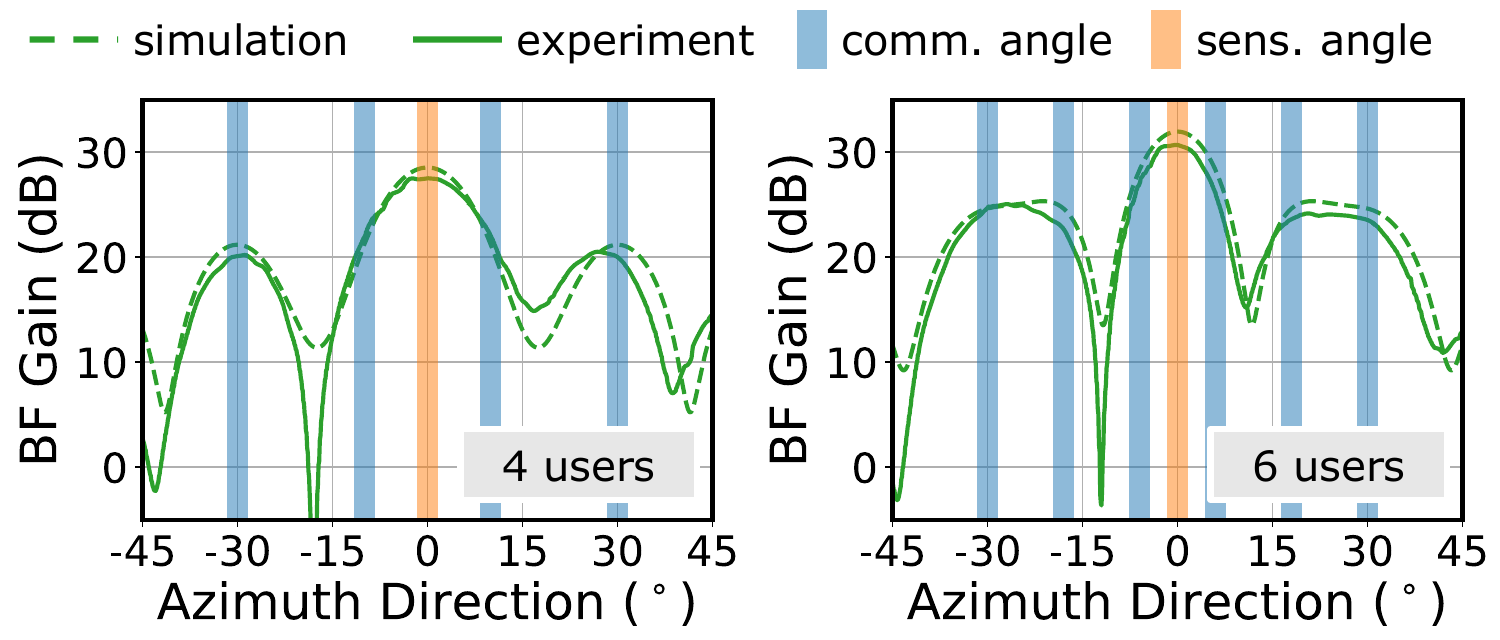}
    \vspace{-3mm}
    \caption{The beamforming pattern with four (left) and six (right) users with sensing angle at 0$^\circ$.}
    \vspace{-3mm}
    \label{fig:beam-pattern-user}
\end{figure}
%% figure ends

\myparatight{Communication user number scalability.}
Finally, we evaluate {\name}'s scalability in scenarios with dense users.
We consider four users evenly located at $\pm 30^{\circ}$ and $\pm10^{\circ}$.
We assume that each user is allocated a {1}$\times${8} subarray, and the whole antenna array dimension during DMRS is scaled up to {4}$\times${8} accordingly.
As shown in Fig.~\ref{fig:beam-pattern-user}(\emph{left}), a wider main lobe is generated by {\BeamOptim}, serving as both the sensing beam at $\angleSenseAz=0^{\circ}$ and the two nearby users at $10^{\circ}$.
With a comparable communication beamforming for the four users ({20.13--22.55}\thinspace{dB}), the sensing beam has a beamforming gain of {27.50}\thinspace{dB}.
Furthermore, for the six-user case where users are evenly located between $\pm 30^{\circ}$, we have a larger whole antenna array as {6}$\times${8}.
In this case, the beamforming gains of the communication beams are between {23.02--27.64}\thinspace{dB}, and the sensing beam has a beamforming gain of {30.68}\thinspace{dB}.
The increased beamforming gain comes from the increased number of antennas in the employed {6}$\times${8} antenna arrays.

\subsection{Communication Performance}
\label{ssec: evaluation-communication}

We evaluate {\name}'s communication capability following the setup in Fig.~\ref{fig:experiment-setup}(\emph{middle}).
Since the two users are symmetric, we only show the communication performance for the user at $+30^{\circ}$.
Note that the SNR reported below is measured during the data symbols.

%% figure begins
\begin{figure}%[!t]
    \centering
    \includegraphics[width=1\linewidth]{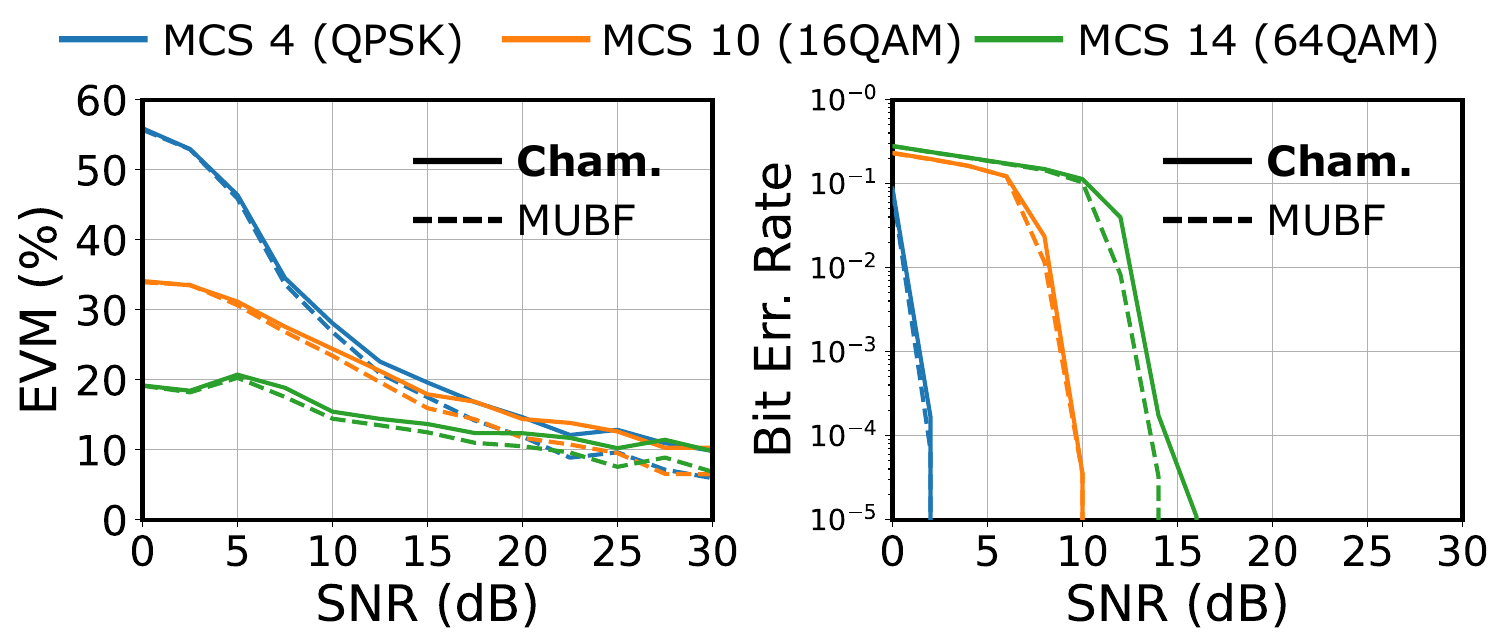}
    \vspace{-3mm}
    \caption{The EVM (\emph{left}) and BER (\emph{right}) comparison of {\namebf} and the MUBF baseline over different SNRs under 5G MCS 4/10/14.}
    \vspace{-3mm}
    \label{fig:communication-metric}
\end{figure}
%% figure ends

\begin{figure}[!t]
    \begin{minipage}{0.49\columnwidth}
        \centering
        \includegraphics[width=1\linewidth]{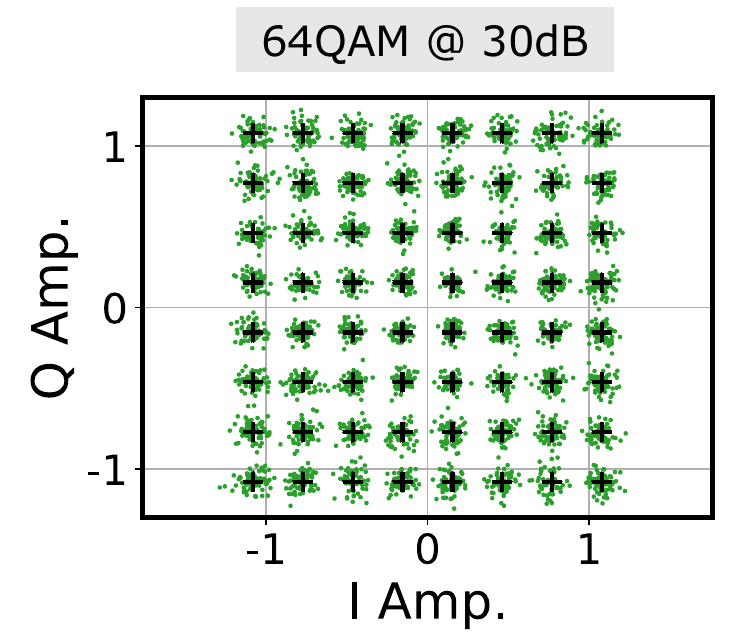}
        \vspace{-3mm}
        \caption{The demodulated constellation by {\namebf} of 64QAM at {30}\thinspace{dB} SNR.}
        \vspace{-3mm}
        \label{fig:communication-constel}
    \end{minipage}
    \hfill
    \begin{minipage}{0.49\columnwidth}
        \centering
        \includegraphics[width=1\linewidth]{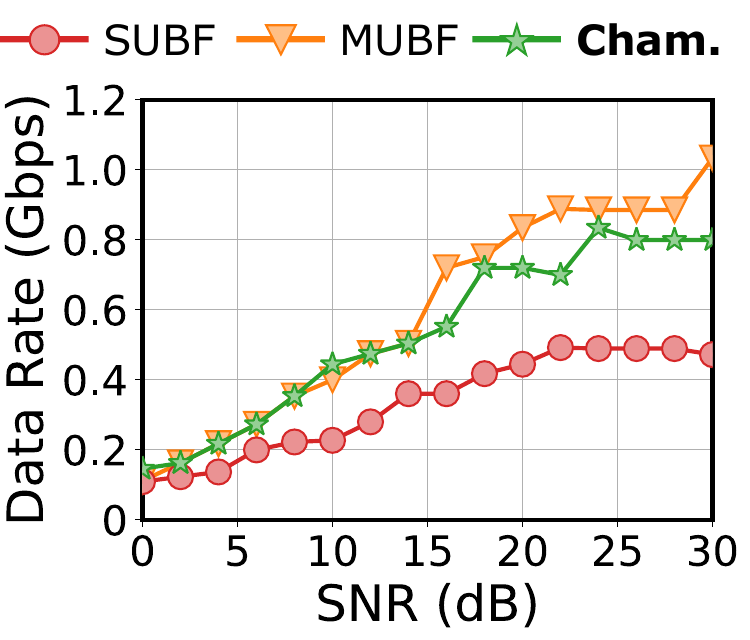}
        \vspace{-3mm}
        \caption{The achieved sum data rate of {\namebf} and the SUBF/MUBF baselines.}
        \vspace{-3mm}
        \label{fig:communication-speed}
    \end{minipage}
\end{figure}

\myparatight{EVM and BER performance.}
We vary the SNR from {0--30}\thinspace{dB} across three 5G MCSs: MCS-4 ({QPSK-$\frac{602}{1,024}$}), MCS-10 ({16QAM-$\frac{658}{1,024}$}), and MCS-14 ({64QAM-$\frac{616}{1,024}$}), corresponding to three modulation schemes with similar code rates.
The EVM comparison between {\name} and the pure communication MUBF baseline, Mambas, is shown in Fig.~\ref{fig:communication-metric}(\emph{left}).
Due to the signal distortion and hardware imperfections in the PAAMs, the EVM of the MUBF baseline approaches {6.82\%} for 64QAM under {30}\thinspace{dB} SNR.
In contrast, {\name} exhibits only modest average EVM degradations of {2.02\%/1.91\%/1.44\%} for the three MCSs.
Specifically for the 64QAM under {30}\thinspace{dB} SNR, Fig.~\ref{fig:communication-constel} shows example constellations over {5} data symbols, demonstrating that with the assistance of {\name}'s pre-distortion at the BS, the communication CSI can still be accurately estimated by the users without significant amplitude or phase errors.
We also evaluate the BER performance with LDPC decoding, as shown in Fig.~\ref{fig:communication-metric}(\emph{right}).
To achieve a BER of $10^{-4}$, {\name} requires SNRs of {4/10/16}\thinspace{dB} for MCS-4/10/14, respectively.
In summary, {\name} achieves communication performance comparable to the MUBF baseline at low SNRs, but reveals a slight performance gap at higher SNRs (e.g., above {20}\thinspace{dB}) due to inaccuracies in communication CSI estimation.

\myparatight{Link data rates.}
Finally, we showcase the sum data rates achieved by {\name} in comparison with the SUBF (one user at a time) and MUBF baselines, as shown in Fig.~\ref{fig:communication-speed}.
Under the SNR of {10/20/30}\thinspace{dB}, the maximum MCSs achieved by {\name} are {9/15/18} while meeting the $<${10}\% BLER requirement, corresponding to a sum data rate across the two users of {0.444/0.719/0.799}\thinspace{Gbps}.
The multi-user support of {\name} has a comparable performance to the MUBF baseline, and effectively boosts the sum data rate by {1.96/1.62/1.69}$\times$ compared to the SUBF baseline, respectively at {10/20/30}\thinspace{dB} SNR, while providing sub-symbol beam switching for sensing.

\subsection{Sensing Performance}
\label{ssec: evaluation-sensing}

%% figure begins
\begin{figure}%[!t]
    \centering
    \includegraphics[width=0.98\columnwidth]{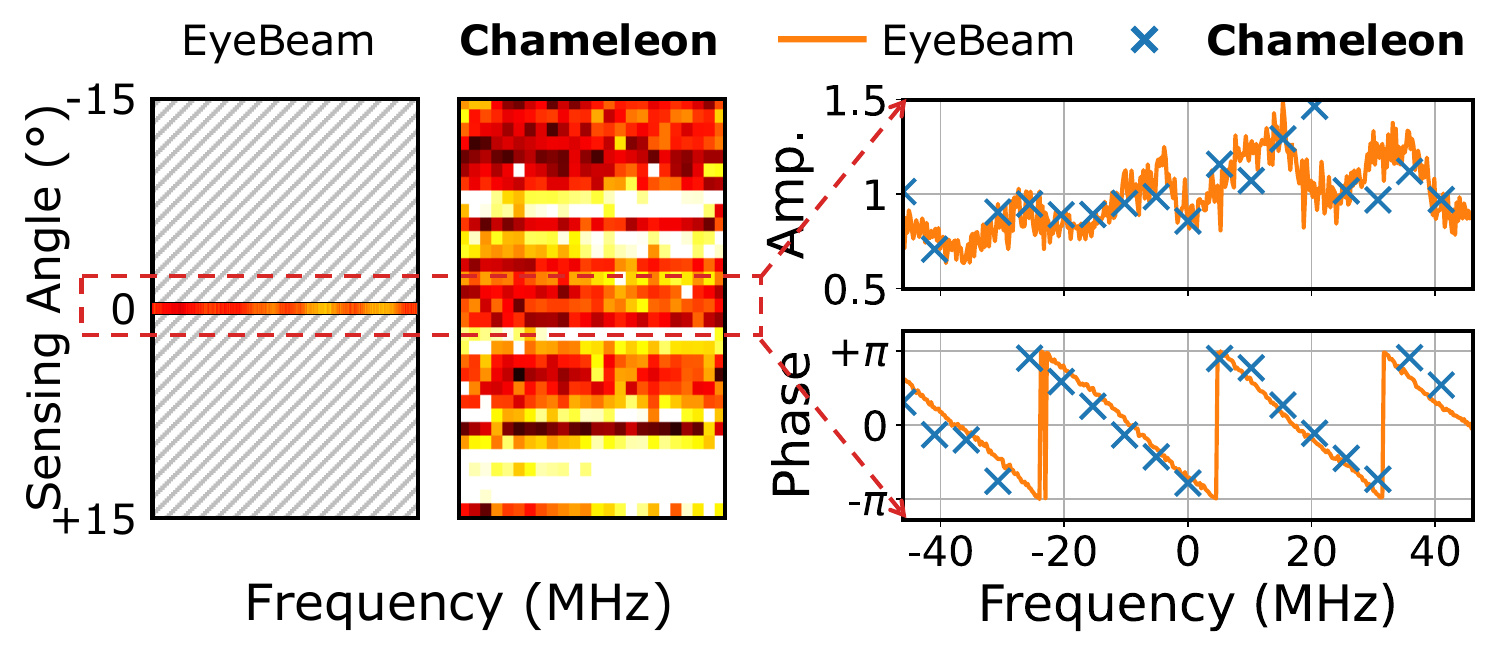}
    \vspace{-3mm}
    \caption{CSI amplitude obtained within one symbol, where {\namebf} provides angular information by beamformer switching (\emph{left}). Specifically for the sensing angle at $0^{\circ}$, the sensing CSI by {\namebf} is close to the full CSI by EyeBeam without switching (\emph{right}).}
    \vspace{-3mm}
    \label{fig:CSI-validate}
\end{figure}
%% figure ends

As shown in Fig.~\ref{fig:experiment-setup}(\emph{right}), we conduct the beamformer switching at an interval of {0.24}\thinspace{\usec} within each DMRS symbol on the TX PAAM, and deploy another RX PAAM to receive the reflected signals and extract the sensing CSI for three applications.
Note that {0.24}\thinspace{\usec}-per-beamformer switching enables {\name} to retrieve sensing CSI {340}$\times$ faster than existing ISAC systems that rely on SSBs.

%%%%%
%%%%%
\myparatight{Sensing CSI validation.}
Fig.~\ref{fig:CSI-validate}(\emph{left}) shows the CSI estimated using a single DRMS symbol of {8.3}\thinspace{\usec}.
With a fixed beamformer, the baseline EyeBeam can only obtain CSI toward a single sensing angle at {0}$^{\circ}$.
In contrast, {\name} with sub-symbol beam switching and the CSI reconstruction algorithm (see \S\ref{ssec: system-design-sensing-csi}) allows CSI estimation across both the angular ({-15--+15}$^{\circ}$) and frequency ({92.16}\thinspace{MHz} bandwidth) domains.
Fig.~\ref{fig:CSI-validate}(\emph{right}) further zooms in on the detailed CSI amplitude and phase responses at the {0}$^{\circ}$ sensing angle.
Because each sensing spans only ({0.24}\thinspace{\usec}), the sensing CSI provided by {\name} has a coarser frequency resolution, while the overall amplitude and phase responses closely match those provided by EyeBeam.

%% figure begins
\begin{figure}%[!t]
    \centering
    \includegraphics[width=0.98\columnwidth]{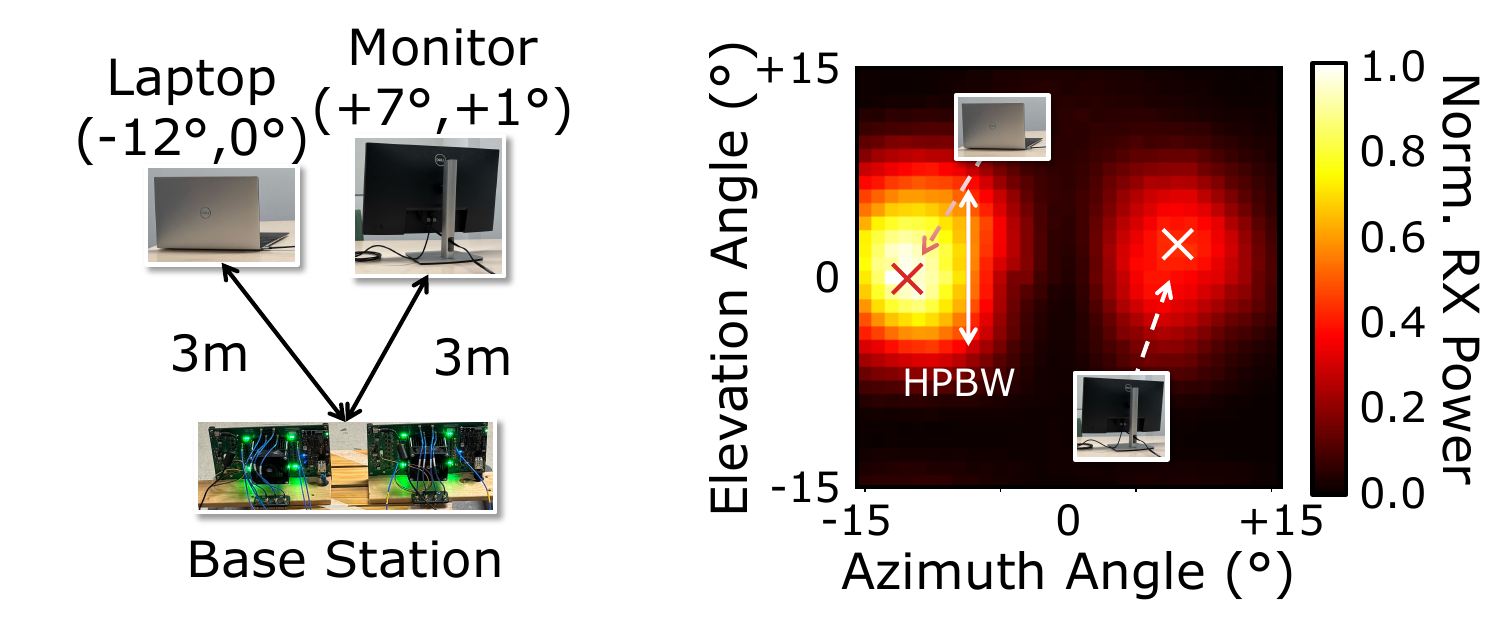}
    \vspace{-3mm}
    \caption{(\emph{Left}) The 2D imaging environment with a laptop and a monitor, and (\emph{Right}) sensing results across azimuth and elevation angles.}
    \vspace{-3mm}
    \label{fig:application-imaging}
\end{figure}
%% figure ends

\myparatight{Application 1: 2D imaging.}
We first demonstrate a 2D imaging application in which the sensing angle sweeps from $-15^{\circ}$ to $+15^{\circ}$ along both azimuth and elevation, resulting in a total of {31}$\times${31} = 961 beamforming directions.
The sensing environment includes a laptop and a monitor as the major reflectors, as illustrated in Fig.~\ref{fig:application-imaging}(\emph{left}).
The corresponding imaging results are shown in Fig.~\ref{fig:application-imaging}(\emph{right}), represented by a {31}$\times${31} heatmap of the overall received power per sensing direction, and two distinct ``spots'' correspond to the positions of the laptop and monitor.
Note that the spot radius is mainly determined by the $12.3^{\circ}$ HPBW of the TX PAAM.
Importantly, the peak azimuth and elevation angles of the two spots accurately localize the two objects, and the reflected power from the laptop is stronger than that from the monitor due to its metallic surface.
With {\name}, scanning across {961} beamformers can be completed within only {8} time slots--equivalent to {0.875}\thinspace{ms} (34 directions per DMRS symbol and four DMRS symbols per slot)--highlighting its potential for fast and large-scale reflection heatmap generation.
The imaging resolution can be proportionally increased with larger antenna arrays.

%% figure begins
\begin{figure}%[!t]
    \centering
    \includegraphics[width=0.98\columnwidth]{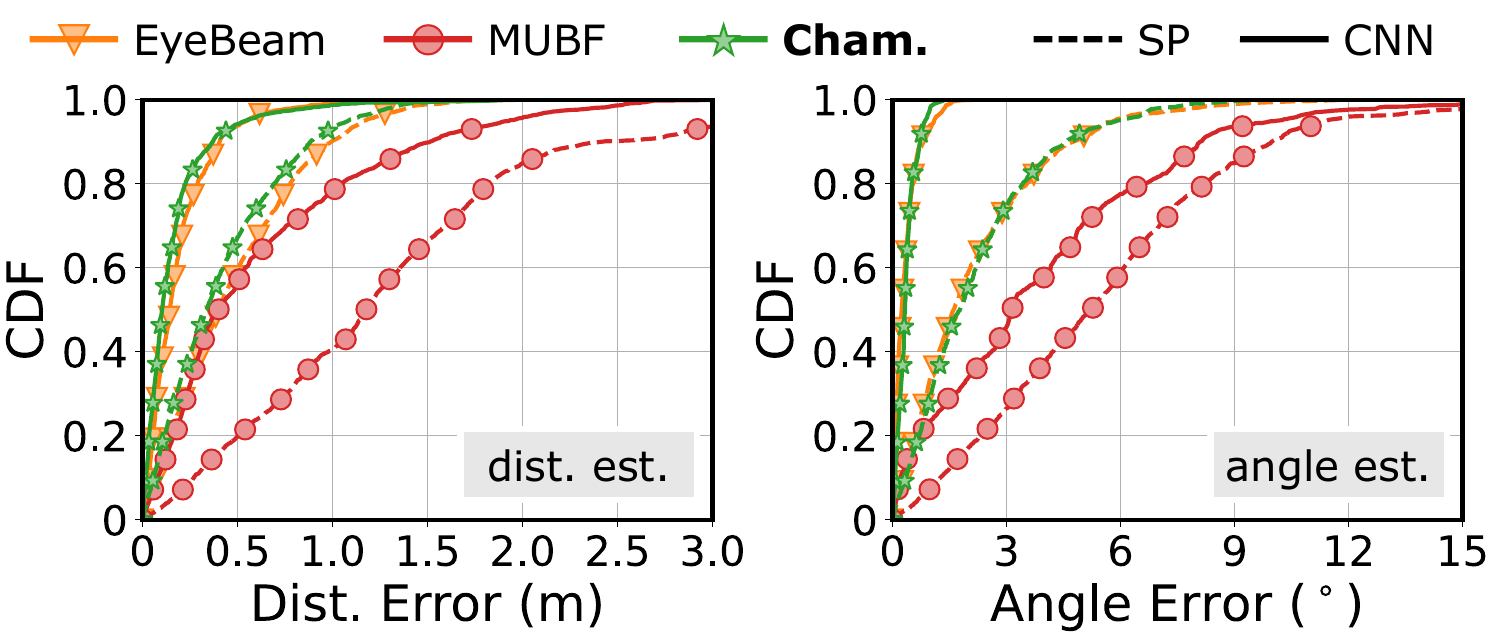}
    \vspace{-3mm}
    \caption{Using the CSI estimated from EyeBeam, MUBF, and {\namebf} with the combination of SP and CNN, the localization error of the metal sheet in terms of the distance (\emph{left}) and angle (\emph{right}).}
    \vspace{-3mm}
    \label{fig:ML-application}
\end{figure}
%% figure ends

\myparatight{Application 2: Object localization.}
In the second application, we apply {\name} to localize the distance and angle of an aluminum metal sheet dimensioned of {0.3}$\times${0.5}\thinspace{m}$^2$.
We extract three features from the CSI estimated from each beamformer: received power, phase slope, and linearity loss, as described in \S\ref{ssec: implementation-system-implementation}.
We first develop a signal processing (SP) method~\cite{gallyas2024angle} that considers a series of pre-defined weights per feature per beamformer; the distance and the angle estimation can be acquired as the weighted sum over the three features extracted from all the CSI.
We also train a convolutional neural network (CNN) with two 1-dimensional convolutional layers backboned by LeNet~\cite{lecun1998gradient} to regress the distance and angle, respectively.
The three features are treated as three channels to the CNN model.
This CNN model is trained by the mean squared error (MSE) loss.
The train and test dataset of the distance estimation is collected from {1}\thinspace{m} to {8}\thinspace{m} every {0.1}\thinspace{m} in front of the TX/RX PAAMs at an angle of {0}$^{\circ}$; for each position, we transmit {25} time slots with a total of {100} DMRS symbols.
As shown in Fig.~\ref{fig:ML-application}(\emph{left}), {\name} achieves median errors of {0.40}\thinspace{m} and {0.14}\thinspace{m} by the SP and CNN methods, respectively, close to the median errors of {0.34}\thinspace{m} and {0.10}\thinspace{m} by EyeBeam.
In addition, we collect a separate dataset for the angle estimation task, where the metal sheet is placed at {3}\thinspace{m} distance within an angle of $\pm${15}$^{\circ}$ at the step of {1}$^{\circ}$. Note that the metal sheet always faces the TX/RX PAAMs when changing the angle.
As shown in Fig.~\ref{fig:ML-application}(\emph{right}), the median errors by {\name} with SP and CNN methods are {1.62}$^{\circ}$ and {0.24}$^{\circ}$, comparable to that achieved by EyeBeam of {1.77}$^{\circ}$ and {0.32}$^{\circ}$, respectively.

%% figure begins
\begin{figure}%[!t]
    \centering
    \vspace{-1mm}
    \includegraphics[width=0.38\columnwidth]{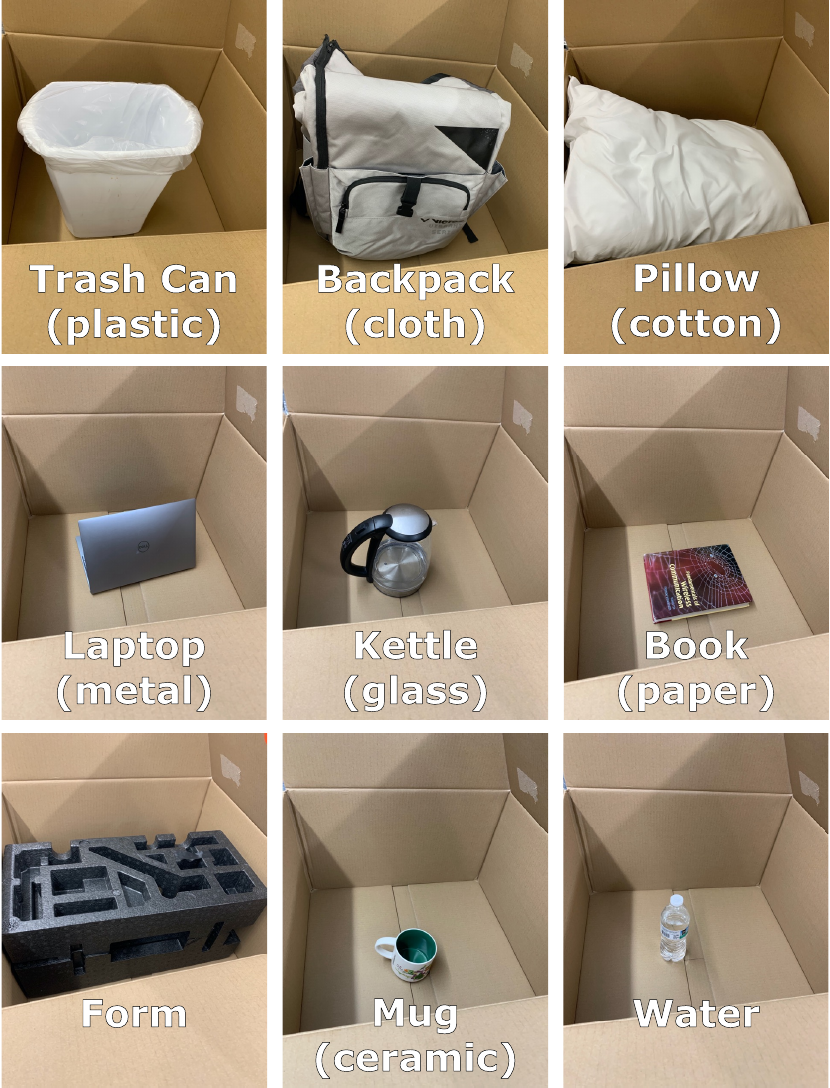}
    \includegraphics[width=0.60\columnwidth]{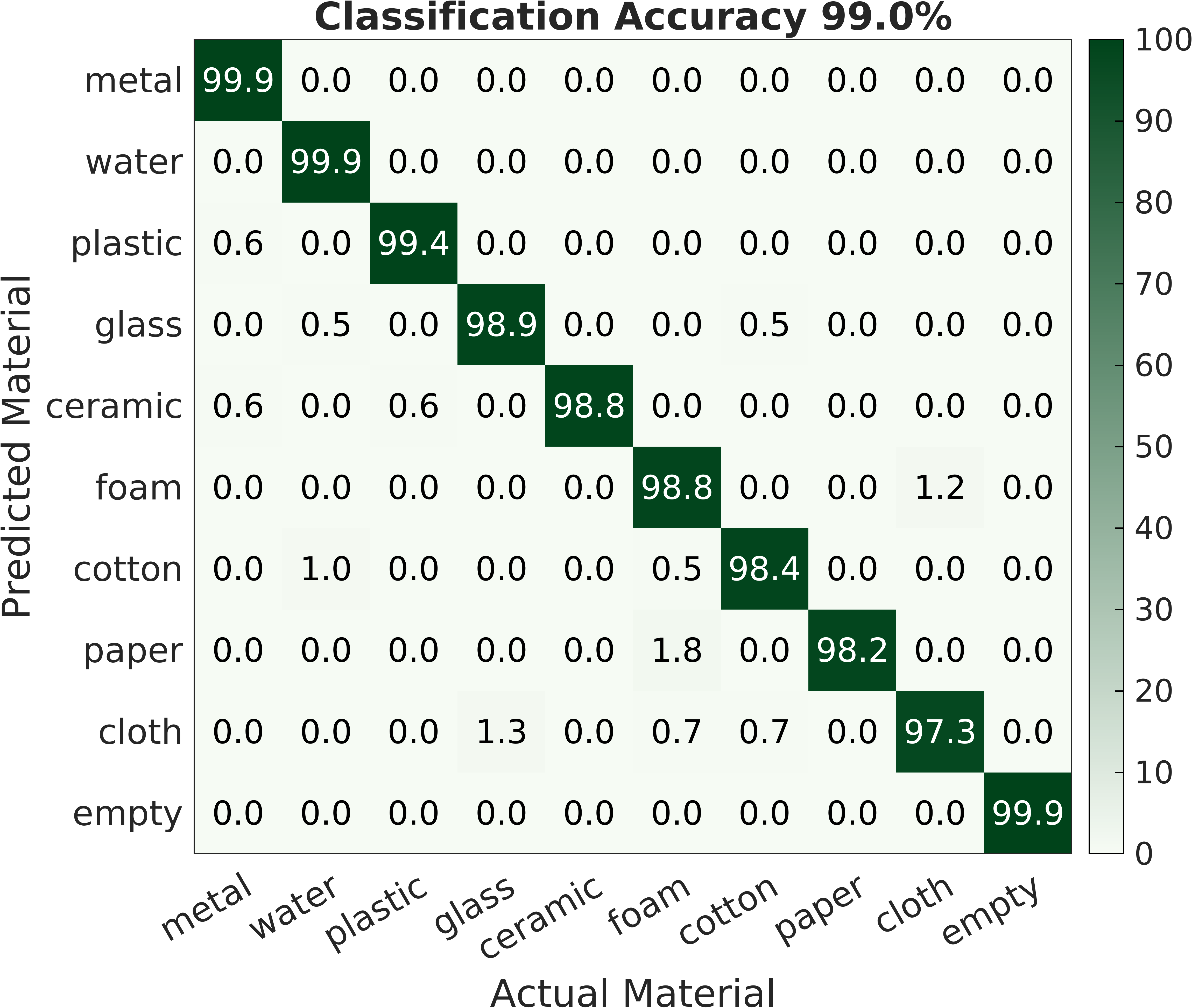}
    \vspace{-3mm}
    \caption{The material classification objects (left) and the classification confusion matrix (right).}
    \vspace{-5mm}
    \label{fig:material-classification}
\end{figure}
%% figure ends

\myparatight{Application 3: Material classification.}
We also consider the application of material classification based on the sensing CSI amplitude due to different impacts on the reflected path (e.g., attenuation and phase shift) when hitting the materials.
Fig.~\ref{fig:material-classification} (\emph{left}) shows the nine objects representing nine materials: the laptop (metal), water, trash can (plastic), kettle (glass), mug (ceramic), foam, pillow (cotton), book (paper), and backpack (cloth), plus an empty case.
We put these objects inside a cardboard in front of the BS of {1}\thinspace{m}.
For each object/material, we collect the dataset of sensing CSI for {100} time slots, including {400} DMRS symbols with {400} sets of sensing CSI.
We train a linear regression model, taking the sensing CSI amplitude as the input to do a 10-classification task.
As a result, this linear regression model achieves a classification accuracy of 99.0\%, and we show the detailed confusion matrix in Fig.~\ref{fig:material-classification}(\emph{right}).

\subsection{Mobility Evaluation}
\label{ssec: evaluation-mobility}

%% figure begins
\begin{figure}%[!t]
    \centering
    \includegraphics[width=0.98\columnwidth]{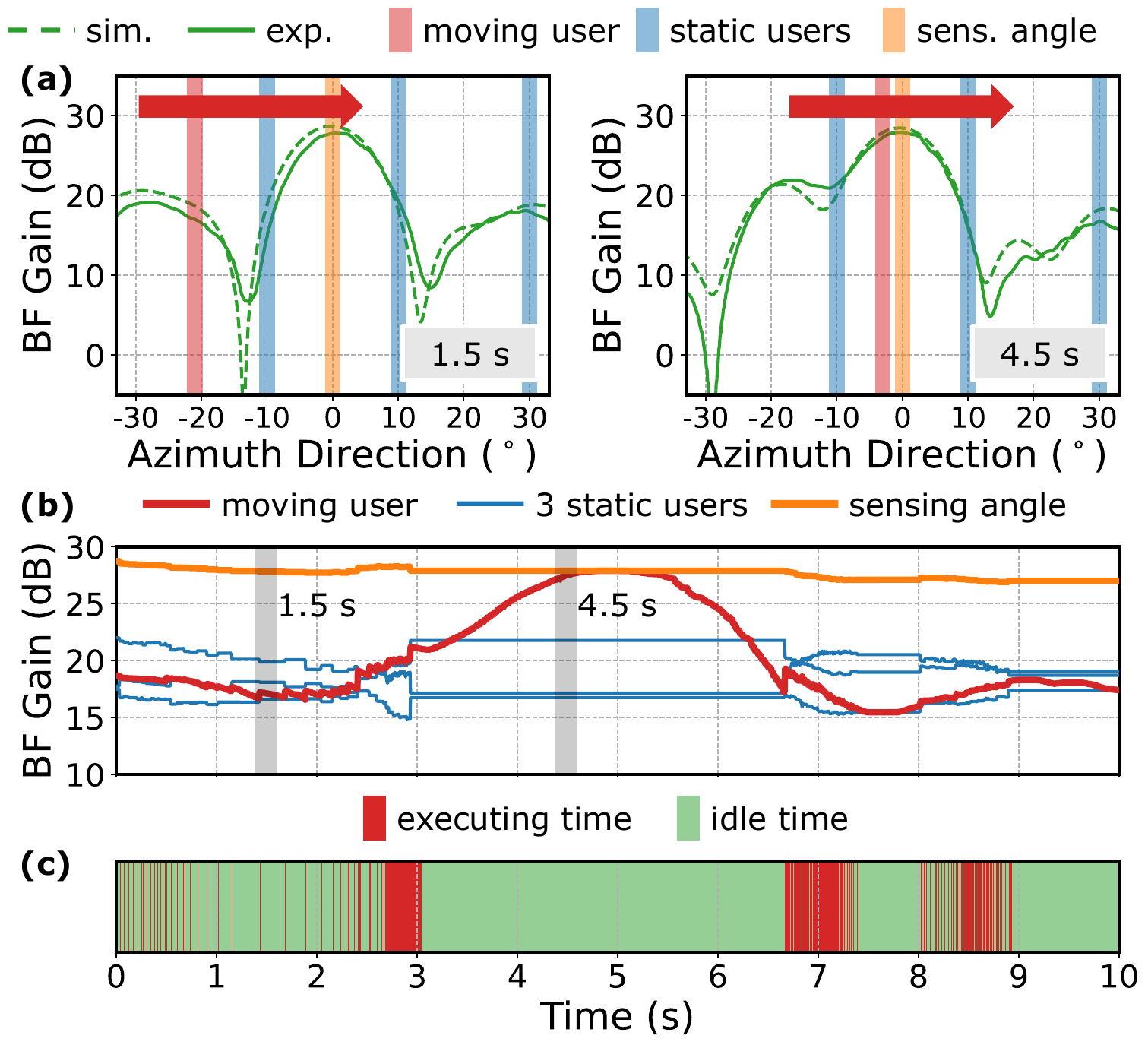}
    \vspace{-3mm}
    \caption{The mobility experiment with three users at {-10/+10/+30}$^{\circ}$ and one moving user from {-30}$^{\circ}$ to {+30}$^{\circ}$ within {10}\thinspace{sec}, where we focus on the beamformer with $\angleSenseAz=0^{\circ}$.
    (a) The beamforming patterns at time stamps at {1.5}\thinspace{sec} (\emph{left}) and {4.5}\thinspace{sec} (\emph{right}).
    (b) The sensing and communication beamforming gain over the {10}\thinspace{sec}.
    (c) On the server, the executing time occupancy to optimize the beamformers by {\BeamOptim}.
    }
    \vspace{-3mm}
    \label{fig:runtime-experiment}
\end{figure}
%% figure ends

We further implement {\name} in an environment with mobile users to evaluate its runtime performance, where the BS employs a {4}$\times${8} array to serve four users.
Initially, the four users are positioned at $\pm 30^{\circ}$ and $\pm10^{\circ}$, with the user at $-30^{\circ}$ moving toward $+30^{\circ}$ within {10}\thinspace{sec}, corresponding to a speed of {0.52}\thinspace{m/s} at a link distance of {5}\thinspace{m}.
For simplicity, we assume a uniform base SNR across all users, such that their effective SNR $\snrComm{\userIdx}$ is proportional to the beamforming gain.
In the meanwhile, {\name} employs Algorithm~\ref{algo:optimize-beam} to optimize the beamformer codebook $\bfWeightVecSet$ every {5}\thinspace{ms}, consistent with the SSB interval used to track user mobility.

\myparatight{Beamforming patterns with moving users.}
Without loss of generality, we showcase the beamformer with the sensing angle at $\angleSenseAz=0^{\circ}$ in Fig.~\ref{fig:runtime-experiment}.
Fig.~\ref{fig:runtime-experiment}(a) shows the beamforming patterns at {1.5}\thinspace{sec} and {4.5}\thinspace{sec}.
At {1.5}\thinspace{sec}, the moving user at $-21^{\circ}$ has a measured beamforming gain of {17.10}\thinspace{dB}, while that of the other three users are {16.58/19.88/18.09}\thinspace{dB}. 
At this time stamp, the moving user is close to the minimum SNR (at a difference of {0.52}\thinspace{dB}), the {\BeamOptim} may be performed frequently.
In this case, this beamformer should be optimized by {\BeamOptim}.
At {4.5}\thinspace{sec}, the user moves to $-3^{\circ}$, close to the sensing angle at $0^{\circ}$.
At this time stamp, its communication beamforming gain is {27.53}\thinspace{dB}, which is not the bottleneck SNR, and the {\BeamOptim} is waived according to Algorithm~\ref{algo:optimize-beam}.

\myparatight{Communication and sensing beamforming gains over time.}
Fig.~\ref{fig:runtime-experiment}(b) further shows the beamforming gains of the four users, and towards the sensing angle at $\angleSenseAz=0^{\circ}$ over the {10}\thinspace{sec} experiment period.
Overall, the mean beamforming gains of the moving users are {20.24}\thinspace{dB}, and those of the three static users are {19.51/18.86/16.90}\thinspace{dB}, respectively.
In addition, the beamforming gain of the sensing angle is maintained between {27.68--29.00}\thinspace{dB}, as designed.

\myparatight{Latency measurement.}
Finally, the optimization latency of {\BeamOptim} is shown in Fig.~\ref{fig:runtime-experiment}(c), measured when Algorithm~\ref{algo:optimize-beam} with {\BeamOptim} runs on a server with a single Intel Xeon 6384 CPU core.
During the {10}\thinspace{sec} experiment, the user angles are updated {2,000} (based on the {5}\thinspace{msec} SSB interval), only {253} out of which {\BeamOptim} is conducted, while others are skipped.
The median executing time to conduct {\BeamOptim} is {1.62}\thinspace{ms}, much lower than the update duration of {5}\thinspace{ms}; the total executing time of {\BeamOptim} is {0.741}\thinspace{sec}, only occupying {7.41\%} of the whole experiment duration of {10}\thinspace{sec}. 
This indicates that a single core can support {14} beamformer updates for {\name} in real-time in dynamic environments with mobile users, and this number can further be improved with more CPU cores.
\section{Conclusions}
\label{sec:conclusion}

In this paper, we presented {\name}, an ISAC system that is compliant with the 5G mmWave networks.
Specifically, we designed a set of ISAC-capable beamformers, each of which has an additional beam toward a desired sensing angle while maintaining the communication beams toward the users. Within a DMRS symbol, the BS switches multiple beamformers and extracts sub-symbol CSI per beamformer for sensing purposes; meanwhile, the users can still estimate the communication CSI for data demodulation.
The proposed {\name} is implemented and evaluated on a {28}\thinspace{GHz} SDR testbed, where sensing CSI can be effectively extracted while incurring subtle communication degradation. Such sensing CSI enables sensing applications, including but not limited to 2D imaging, object localization, and material classification.

\begin{acks}
This work was supported in part by NSF grants CNS-2112562, CNS-2211944, AST-2232458, and CNS-2443137.
This work was also supported in part by the Center for Ubiquitous Connectivity (CUbiC), sponsored by Semiconductor Research Corporation (SRC) and Defense Advanced Research Projects Agency (DARPA) under the JUMP 2.0 program.
We thank Arun Paidimarri, Asaf Tzadok, Alberto Valdes-Garcia (IBM Research), Jakub Kolodziejski, Prasanthi Maddala, and Ivan Seskar (Rutgers University) for their contributions to this work.
\end{acks}

\bibliographystyle{ACM-Reference-Format}
\bibliography{reference}

\end{document}